\documentclass[conference]{IEEEtran}
\IEEEoverridecommandlockouts
\usepackage{xspace}
\usepackage{balance}  % for  \balance command ON LAST PAGE  (only there!)
\usepackage{mathtools}
\usepackage{mdframed}
\usepackage{array}
\usepackage{textcomp}
\usepackage{xcolor}
\usepackage{framed}
\usepackage{bm}
\usepackage[mathscr]{euscript}
\usepackage{subfigure, multirow}
\usepackage{xspace}
\usepackage{booktabs}
\usepackage{times,epsfig,epsf,psfrag,array,amsmath,amssymb,graphicx,graphics,verbatim}
\usepackage[ruled,vlined,linesnumbered, commentsnumbered]{algorithm2e}
\usepackage{url}
\usepackage{cite}
\usepackage{comment}
\usepackage{enumerate}
\usepackage{float}

\newfloat{figtab}{htb}{fgtb}
\makeatletter
\newcommand\figcaption{\def\@captype{figure}\caption}
\newcommand\tabcaption{\def\@captype{table}\caption}
\makeatother

\newtheorem{definition}{Definition}
\newtheorem{theorem}{Theorem}
\newtheorem{lemma}{Lemma}

\newtheorem{example}{Example}

\newcommand{\eat}[1]{}
\newcommand{\eg}{{\textrm{e.g.}}\xspace}

\newcommand{\ie}{{\textrm{i.e.}}\xspace}
\newcommand{\etal}{{\textrm{et al.}}\xspace}

\newcommand{\cS}{\mathcal{C}\xspace}

\SetKwFunction{Simple}{\sc Simple}
\SetKwFunction{Swap}{\sc Swap}
\SetKwFunction{OneSwap}{\sc OneSwap}
\SetKwFunction{TwoSwap}{\sc TwoSwap}
\SetKwFunction{Find}{\sc Update-$\cS$}
\SetKwFunction{MoveIn}{\sc MoveIn}
\SetKwFunction{MoveOut}{\sc MoveOut}

\def\BibTeX{{\rm B\kern-.05em{\sc i\kern-.025em b}\kern-.08em
		T\kern-.1667em\lower.7ex\hbox{E}\kern-.125emX}}

\begin{document}
	%\input{response}
	%\clearpage

	\title{Dynamic Approximate Maximum Independent Set on Massive Graphs}
	
	\author{\IEEEauthorblockN{Xiangyu Gao$^{\hspace{.1em}\S\ddag}$, Jianzhong Li$^{\hspace{.1em}\ddag\S}$, Dongjing Miao$^{\hspace{.1em}\S}$}
		\vspace{.4em}
		\IEEEauthorblockA{\textit{$^\S$Department of Computer Science and Technology, Harbin Institute of Technology, Harbin, China} \\
			\textit{$^\ddag$Faculty of Computer Science and Control Engineering, Shenzhen Institute of Advanced Technology} \\ 
			\textit{Chinese Academy of Sciences, Shenzhen, China} \\
			\vspace{4pt}
			gaoxy@hit.edu.cn, lijzh@siat.ac.cn, miaodongjing@hit.edu.cn}
	}
	\maketitle

	\begin{abstract}
Computing a maximum independent set (MaxIS) is a fundamental NP-hard problem in graph theory, which has important applications in a wide spectrum of fields.
Since graphs in many applications are changing frequently over time, the problem of maintaining a MaxIS over dynamic graphs has attracted increasing attention over the past few years.
Due to the intractability of maintaining an exact MaxIS, this paper aims to develop efficient algorithms that can maintain an approximate MaxIS with an accuracy guarantee theoretically.
In particular, we propose a framework that maintains a $(\frac{\Delta}{2} + 1)$-approximate MaxIS over dynamic graphs and prove that it achieves a constant approximation ratio in many real-world networks.
To the best of our knowledge, this is the first non-trivial approximability result for the dynamic MaxIS problem.
Following the framework, we implement an efficient linear-time dynamic algorithm and a more effective dynamic algorithm with near-linear expected time complexity.
Our thorough experiments over real and synthetic graphs demonstrate the effectiveness and efficiency of the proposed algorithms, especially when the graph is highly dynamic.
\end{abstract}
	\section{Introduction}\label{sec:intro}

\begin{figure}[h]\vspace{-2mm}
	\centering
	\subfigure[Maximal independent set.]{
		\includegraphics[width=0.45\linewidth]{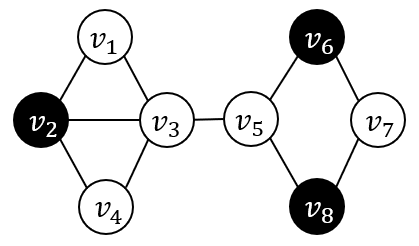}
		\label{fig:maximalIS}
	}
	\subfigure[Maximum independent set.]{
		\includegraphics[width=0.45\linewidth]{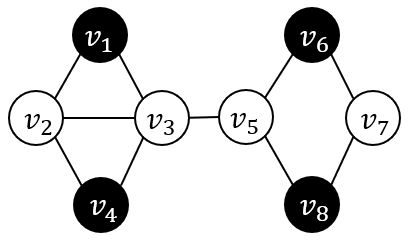}
		\label{fig:maximumIS}
	}\vspace{-1ex}
	\caption{An example graph to illustrate independent sets.}
	\label{fig:IS}
	\vspace{-2mm}
\end{figure}

Graph has been used to model many types of relationships among entities in a wide spectrum of applications such as bioinformatics, semantic web, social networks, and software engineering.
Significant research efforts have been devoted towards many fundamental problems in managing and analyzing graph data.
The {\it maximum independent set} (MaxIS) problem is a classic NP-hard problem in graph theory~\cite{DBLP:books/fm/GareyJ79}.
Given a graph $G$, a subset $I$ of vertices in $G$ is an \textit{independent set} if there is no edge between any two vertices in $I$.
A \textit{maximal independent set} is an independent set such that adding any other vertex to the set forces it to contain an edge.
The independent set with the largest size, measured by the number of vertices in it, among all independent sets in $G$ is called the \textit{maximum independent set} in $G$, which may not be unique.
For example, in Fig.~\ref{fig:IS}, $\{v_2, v_6, v_8\}$ is a maximal independent set of size 3, while both $\{v_1, v_4, v_6, v_8\}$ and $\{v_1, v_4, v_5, v_7\}$ are maximum independent sets of size 4.

The MaxIS problem has a lot of real-world applications, such as indexing techniques ~\cite{DBLP:journals/pvldb/FuWCW13, DBLP:journals/pvldb/JiangFWX14}, collusion detection~\cite{DBLP:journals/jpdc/AraujoFDSK11, DBLP:journals/pvldb/MiaoCLGL20, DBLP:journals/tcs/MiaoLLL19}, automated map labeling~\cite{DBLP:conf/wea/GemsaNR14}, social network analysis~\cite{DBLP:conf/wea/GoldbergHM05}, and association rule mining~\cite{DBLP:conf/kdd/ZakiPOL97}.
Additionally, it is also closely related to a series of well-known graph problems, such as \textit{minimum vertex cover}, \textit{maximum clique}, and {\it graph coloring}.
%It is easy to see that finding a maximum clique of a graph $G$ is equivalent to finding a MaxIS of the complement of $G$.
%And, to find a minimum vertex cover of $G = (V, E)$, it is sufficient to compute a MaxIS $I$ of $G$ and return $V\setminus I$ as the result.
Because of its importance, the MaxIS problem has been extensively studied for decades.
Since it is NP-hard to find a MaxIS, the worst-case time complexities of all known exact algorithms are exponential in $n$, the number of vertices in the graph.
The worst-case time complexity of the state-of-the-art exact algorithm is $O(1.1996^nn^{O(1)})$~\cite{DBLP:journals/iandc/XiaoN17}, which is obviously unaffordable in large graphs.
Moreover, the MaxIS problem is also hard to be approximated.
It is proved that the MaxIS problem can not be approximated within a constant factor on general graphs~\cite{DBLP:journals/jal/Robson86}, and for any $\varepsilon \in (0, 1)$, there is no polynomial-time $n^{1-\varepsilon}$-approximation algorithm for it, unless NP = ZPP~\cite{DBLP:conf/focs/Hastad96}.
As a result, the approximation ratios of the existing methods depend on either $n$ or $\Delta$, where $\Delta$ is the maximum degree of $G$.
Till now, the best approximation ratio known for the MaxIS problem is $O(n(\log{\log{n}})^2/(\log{n})^3)$~\cite{DBLP:journals/siamdm/Feige04}.
In recent years, a lot of research has been devoted to efficiently computing a \textit{near-maximum} (maximal and as large as possible) independent set~\cite{DBLP:journals/heuristics/AndradeRW12, DBLP:conf/sigmod/ChangLZ17, DBLP:conf/wea/DahlumLS0SW16, DBLP:journals/heuristics/GrossoLP08, DBLP:conf/alenex/LammS0SW16, DBLP:journals/pvldb/LiuLYXW15}.
The latest method is proposed by Chang \etal~\cite{DBLP:conf/sigmod/ChangLZ17}, which iteratively applies exact and inexact reduction rules on vertices until the graph is empty.

Although the existing methods are quite efficient and effective, they essentially assume that the graph is static.
However, graphs in many real-world applications are changing continuously, where vertices/edges are inserted/removed dynamically.
For instance, the users in a social network may add new friends or remove existing friendships, and new links are constantly established in the web due to the creation of new pages.
Given such dynamics in graphs, the existing approaches need to recompute the solution from scratch after each update, which is obviously time consuming, especially in large-scale frequently updated graphs.
Therefore, the problem of maintaining a MaxIS over dynamic graphs has received increasing attention over the past few years.

Zheng~\etal~\cite{DBLP:conf/icde/ZhengWYC018} are the first to study the maintenance of a MaxIS over dynamic graphs.
They prove that it is NP-hard to maintain an exact MaxIS over dynamic graphs, and design a lazy search strategy to enable the maintenance of a near-maximum independent set.
However, when the initial independent set is not optimal, the quality of the maintained solution is not satisfying after a few rounds of updates.
To overcome this shortcoming, Zheng \etal~\cite{DBLP:conf/icde/ZhengPCY19} propose an index-based framework.
When a set of vertices is moved out of the current solution, the algorithm looks for a set of complementary vertices of at least the same size based on the index to avoid the degradation of the solution quality.
Experimental results show that their method is less sensitive to the quality of the initial independent set, and is efficient and effective when the number of updates is small.
Whereas, it is observed that the structures of many real-world networks like Facebook and Twitter are highly dynamic over time.
For example, the amounts of reads and comments on some hot topics may grow to more than a million in few minutes, which is almost equal to the number of vertices in the graph.
In this scenario, the complementary relation between vertices represented by the index could become quite complicated, which results in an excessive long search time in their algorithm.
And it is also expensive to ensure the efficiency by restarting their method frequently.
Moreover, none of the existing algorithms provides an accuracy guarantee theoretically.
As the graph evolves, the quality of the solution may drop dramatically.

To address the above issues, this paper studies the problem of maintaining an approximate MaxIS with a non-trivial theoretical accuracy guarantee (less than $\Delta + 1$) over dynamic graphs. 
Instead of finding a set of complementary vertices globally, we resort to the local swap operation which has been shown to be effective in improving the quality of a resultant independent set in static graphs~\cite{DBLP:journals/heuristics/AndradeRW12, DBLP:journals/pvldb/LiuLYXW15}.
However, there are still two major challenges under the dynamic setting.
Firstly, none of the existing work makes a thorough quantitative analysis of how much this strategy could benefit.
The authors of~\cite{DBLP:journals/pvldb/LiuLYXW15} only derive an expected lower bound on the solution size under the power-law random graph model~\cite{DBLP:conf/stoc/AielloCL00}.
However, this model is too strict to describe dynamic graphs as it assumes the amount of vertices with a certain degree to be an exact number.
And their analysis heavily relies on the greedy algorithm used for the initial independent set, which no longer holds when the graph is dynamically updated.
In this paper, we introduce a graph partitioning strategy, and derive a deterministic lower bound on the solution size by considering its projection in each component individually.
We show an optimal case for 1-swap, and prove that the lower bound will not be better by considering more kinds of swaps.
This indicates the limitation of all swap-based approaches to the MaxIS problem.
Moreover, we obtain a more useful lower bound on the solution size in a majority of real-world based on the power-law bounded graph model~\cite{DBLP:journals/algorithmica/ChauhanFR20}.

Secondly, we need a sound and complete schema to ensure that all valid swaps can be identified efficiently after each update.
We propose a framework for maintaining an independent set without $j$-swaps for all $j \le k$, where $k$ is a user-specified parameter balancing the solution quality and the time consumption.
In the framework, we design an efficiently updatable hierarchical structure for storing the information needed for identifying swaps, and find swaps in a bottom-up manner among all candidates to reduce the search space.
Several optimization strategies are also devised to further improve the performance.
Following the framework, we instantiate an efficient linear-time dynamic algorithm and a more effective dynamic algorithm with near linear expected time complexity in power-law bounded graphs.

\noindent{\bf Contributions.} The main contributions of this paper are summarized as follows.
\begin{enumerate}[$\bullet$]
\item We propose a framework that maintains a $k$-maximal independent set over dynamic graphs.
The approximation ratio achieved by it is $\frac{\Delta}{2} + 1$ in general graphs,  and a parameter-dependent constant in power-law bounded graphs.
\item We implement a linear time dynamic $(\frac{\Delta}{2} + 1)$-approximation algorithm by setting $k = 1$.
To the best of our knowledge, this is the first algorithm for the dynamic MaxIS problem with a non-trivial approximation ratio.
\item To further improve the quality of the solution, we implement a near-linear time dynamic $(\frac{\Delta}{2} + 1)$-approximation algorithm by setting $k = 2$.
Experiments show that it indeed maintains a better solution with little time increase.
\item We conduct extensive experiments over a bunch of large-scale graphs.
As confirmed in the experiments, the proposed algorithms are more effective and efficient than state-of-the-art methods, especially when the number of updates is huge.
\end{enumerate}

The reminder of this paper is organized as follows.
Preliminaries are introduced in Section~\ref{sec:pre}.
The framework is presented in Section~\ref{sec:gen-swap}.
Two concrete dynamic algorithms are instantiated in Section~\ref{sec:oneswap}.
Experimental results are reported in Section~\ref{sec:exp}, and the paper is finally concluded in Section~\ref{sec:con}.
%The related work is surveyed in Section~\ref{sec:relatedwork},

	\section{Preliminaries}\label{sec:pre}

In this section, we introduce some basic notations and formally define the problem studied in this paper.

We focus on \textit{unweighted undirected} graphs, and refer them as graphs for ease of representation.
A \textit{dynamic graph} $\mathcal{G}$ is a graph sequence $\langle G_0, \cdots, G_t, G_{t+1}, \cdots \rangle$, where each graph $G_t$ is obtained from its preceding graph $G_t$ by either inserting/deleting a(n) vertex/edge.
For each graph $G_t = (V_t, E_t)$, let $n_t = |V_t|$ and $m_t = |E_t|$ denote the number of vertices and edges in it, respectively.
The \textit{open neighborhood} of a vertex $v$ in $G_t$ is defined as $N_t(v) = \{u \in V_t \mid (u, v) \in E_t\}$, and the \textit{degree} of $v$ is defined as $d_t(v) = |N_t(v)|$.
And the {\it closed neighborhood} of $v$ is defined as $N_t[v] = N_t(v) \cup \{v\}$.
Analogously, given a vertex set $S \subseteq V_t$, the open and closed neighborhood of $S$ is denoted by $N_t(S) = \{v \in V_t \setminus S \mid \exists u \in S : (u, v) \in E_t\}$ and $N_t[S] = N_t(S) \cup S$, respectively.
And let $G_t[S]$ denote the subgraph of $G_t$ induced by $S$.
%And given an integer $k \ge 1$, let $[k]$ denote the integer set $\{1, \cdots, k\}$.

\begin{definition}[Independent Set]
Given a graph $G$, a vertex subset $I$ is an independent set of $G$ if for any two vertices $u$ and $v$ in $I$, there is no edge between $u$ and $v$ in $G$.
\end{definition}

An independent set $I$ is a \textit{maximal independent set} if there does not exist a superset $I'$ of $I$ such that $I'$ is also an independent set.
The \textit{size} of $I$, denoted by $|I|$, is defined as the number of vertices in it.
A maximal independent set $I$ in $G$ is a \textit{maximum independent set} if its size is the largest among all independent sets in $G$, and this size is called the \textit{independence number} of $G$, denoted by $\alpha(G)$.
We say that an independent set $I$ is a $r$\textit{-approximate maximum independent set} in $G$ if ${\alpha(G)} \le r \cdot |I|$, and $I$ is a $r$-approximate maximum independent set over a dynamic graph $\mathcal{G}$ if $I$ remains a $r$-approximate maximum independent in each $G_t \in \mathcal{G}$.

\noindent{\bf Problem Statement.}
Given a dynamic graph $\mathcal{G}$, the problem studied in this paper is to efficiently maintain an $r$-approximate maximum independent set $I$ over $\mathcal{G}$ such that $r < \Delta + 1$.

As proved in~\cite{DBLP:conf/icde/ZhengWYC018}, it is NP-hard to maintain an exact MaxIS over dynamic graphs.
Similarly, the following theorem can be derived directly from the hardness result shown in~\cite{DBLP:conf/focs/Hastad96}.

\begin{theorem}\label{thm:unapprox-lower-bounds}
	Given a dynamic graph $\mathcal{G}$, there is no algorithm that can maintain a $n^{1 - \varepsilon}$-approximate maximum independent set over $\mathcal{G}$ in polynomialt time for any constant $\varepsilon \in (0, 1)$, unless NP = ZPP.
\end{theorem}
\begin{IEEEproof}
	Given a graph $G = (V, E)$, we construct a $(m + 1)$-length dynamic graph $\mathcal{G} = \langle G_0, \cdots, G_m\rangle$  as follows, where $m$ is the number of edges in $G$.
	Let $G_0 = (V, \emptyset)$.
	And for each $i \in \{1, \cdots, m\}$, $G_{i}$ is obtained by inserting an edge of $E \setminus E_{i - 1}$ to $G_{i - 1}$.
	It is easy to see that $G_m = G = (V, E)$.
	Supposing that a $n^{1 - \varepsilon}$-approximate independent set for any constant $\varepsilon \in (0, 1)$ can be maintained in polynomial time over $\mathcal{G}$, then such a good approximation result in $G$ can also be computed in polynomial time, which contradicts to the hardness result shown in~\cite{DBLP:conf/focs/Hastad96}.
\end{IEEEproof}

	\section{A Framework For Maintenance}\label{sec:gen-swap}

In this section, we first analyze the lower bound on the size of $k$-maximal independent sets, and then introduce a framework that efficiently maintains a $k$-maximal independent set over dynamic graphs.

\subsection{$k$-maximal Independent Set}\label{sec:lower-bound}

Given a graph $G$ and an independent set $I$ in $G$, a \textit{$k$-swap} consists of removing $k$ vertices from $I$ and inserting at least $k + 1$ vertices into it.
We say that an independent set $I$ is {\it $k$-maximal} if there is no $j$-swap available in $I$ for all $j \in [k]$, where $[k]$ denotes the set of integers $\{1, \cdots, k\}$.
In what follows, suppose that $I$ is a $k$-maximal independent set, and let $\bar{I} = V \setminus I$.
Notice that $\bar{I}$ can be partitioned into $\Delta$ disjoint subsets $\bar{I}_1, \cdots, \bar{I}_\Delta$, where $\bar{I}_j = \{v \in \bar{I} \mid \vert N(v) \cap I \vert = j\}$ and $\Delta$ is the maximum degree of $G$.
It is easy to see that
\begin{equation}\label{eq:sum_n}
	\begin{small}
		\alpha(G) \le n	= |I| + \lvert \bar{I}_1\rvert + \lvert \bar{I}_2 \rvert + \cdots + \lvert \bar{I}_\Delta\rvert,
	\end{small}
\end{equation}
where $\alpha(G)$ is the independence number of $G$.
Since there is no edge between any two vertices in $I$, it is also derived that
\begin{equation}\label{eq:deg_s}
	\begin{small}
		|\bar{I}_1| + 2|\bar{I}_2| + \cdots + \Delta |\bar{I}_\Delta| = \sum_{v \in I} d(v) \le \Delta\cdot |I|.
	\end{small}\vspace{-1ex}
\end{equation}

\begin{figure}[htb]\vspace{-2ex}
	\centering
	\includegraphics[width=.6\linewidth]{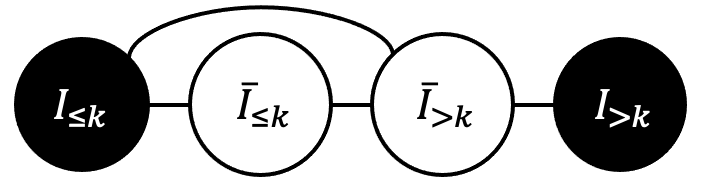}
	\vspace{-1ex}
	\caption{An illustration for graph partition.}
	\label{fig:graph-partition}
	\vspace{-4mm}
\end{figure}

Let $I_{opt}$ denote a MaxIS in $G$ and recall that $|I_{opt}| = \alpha(G)$.
To derive the lower bound on $|I|$, we partition $G$ into two components, and quantify the relationship between $|I|$ and $|I_{opt}|$ in each component separately.
Let $\bar{I}_{\le k} = \cup_{j \in [k]} \bar{I}_j$ and $I_{\le k} = \{v \in I \mid N(v) \cap \bar{I}_{\le k} \ne \emptyset\}$.
Notice that any possible $j$-swap for $j \in [k]$ can only appear in $G[I_{\le k} \cup \bar{I}_{\le k}]$.
And let $\bar{I}_{> k} = \bar{I} \setminus \bar{I}_{\le k}$ and $I_{> k} = I \setminus I_{\le k}$ denote the set of remaining vertices of $\bar{I}$ and $I$, respectively.
These various sets are depicted schematically in Fig.~\ref{fig:graph-partition}.

First, the size of the projection of $I_{opt}$ on $G[I_{> k} \cup \bar{I}_{> k}]$ can not exceed the number of vertices in the subgraph.
Discard the first $k$ items of equation~\ref{eq:deg_s}, it is derived that
\begin{equation}\label{eq:m-g-k}
	%\begin{small}
		\begin{aligned}
			|I_{opt} \cap (I_{>k} \cup \bar{I}_{>k})|& \le |\bar{I}_{k + 1}| + \cdots + |\bar{I}_\Delta| + |I_{> k}|\\
			&\le \frac{\Delta}{k + 1}\cdot |I| + |I_{>k}|.
		\end{aligned}
	%\end{small}
\end{equation}
Then, since $I_{opt} \cap (I_{\le k} \cup \bar{I}_{\le k})$ remains a valid independent set in $G[I_{\le k} \cup \bar{I}_{\le k}]$, we utilize the fact that $I_{\le k}$ is also a $k$-maximal independent set in $G[I_{\le k} \cup \bar{I}_{\le k}]$ to derive an upper bound on $|I_{opt} \cap (I_{\le k} \cup \bar{I}_{\le k})|$.
The following lemma shows an optimal case when $k = 1$.
\begin{lemma}\label{lma:M1-opt}
	Suppose $I$ is an 1-maximal independent set, then $I_1$ is a maximum independent set in $G[I_1 \cup \bar{I}_1]$.
\end{lemma}
\begin{IEEEproof}
	Since there is no 1-swap in $I$, for each vertex $v \in I$, the subgraph induced by $N(v) \cap \bar{I}_1$ is a complete graph.
	For contradiction, suppose that $I'_1$ is a MaxIS of $G[I_1 \cup \bar{I}_1]$ and $|I'_1| > |I_1|$.
	Since $I_1 \nsubseteq I'_1$, $|I'_1\setminus I_1| \ge |I_1 \setminus I'_1| + 1$.
	Due to the Pigeonhole Principle, there must exists at least one vertex $v \in I_1$ having two non-adjacent neighbors in $\bar{I}_1$.
	This contradicts to the fact that $G[N(v) \cap \bar{I}_1]$ is a complete graph for each vertex $v \in I_1$.
	Thus, $I_1$ is a MaxIS of $G[I_1 \cup \bar{I}_1]$.
\end{IEEEproof}
Combining things together, the following theorem is obtained.
\begin{theorem}\label{thm:oneswap-approx-ratio}
	If $I$ is an 1-maximal independent set in $G$, then $\alpha(G) \le (\frac{\Delta}{2} + 1)|I|$.
\end{theorem}
\begin{IEEEproof}
	Since $I_1$ is a MaxIS of $G[I_1 \cup \bar{I}_1]$ and the projection of $I_{opt}$ remains a valid independent set in $G[I_1 \cup \bar{I}_1]$, it is known that $|I_{opt} \cap (I_1 \cup \bar{I}_1)| \le |I_1|$.
	Combining with equation~\ref{eq:m-g-k}, it is derived that
	\begin{displaymath}
		%\begin{small}
			\begin{aligned}
				\alpha(G) = |I_{opt}| &= |I_{opt}\cap (I_1 \cup \bar{I}_1)| + |I_{opt} \cap (I_{>1} \cup \bar{I}_{>1})|\\
				&\le |I_1| + |\bar{I}_{2}| + \cdots |\bar{I}_{\Delta}| + |I_{> 1}|\\
				&\le |I| + \frac{\Delta}{2}\cdot |I|.
			\end{aligned}
		%\end{small}
	\end{displaymath}
\end{IEEEproof}

Counter-intuitively, the following theorem indicates that the lower bound will not be better by considering a larger $k$, \ie, allowing more kinds of swaps.

\begin{figure}[htb]\vspace{-4mm}
	\centering
	\subfigure[Worst case for $k = 3$.]{
		\includegraphics[width=0.32\linewidth]{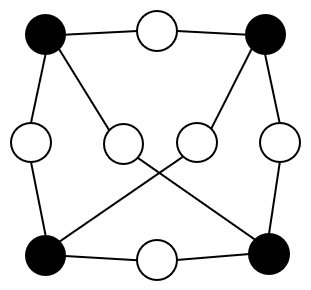}
		\label{fig:worst-case-1}
	}\hspace{2ex}
	\subfigure[Worst case for $k = 4$.]{
		\includegraphics[width=0.32\linewidth]{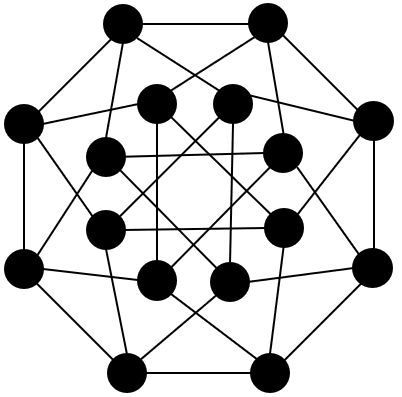}
		\label{fig:worst-case-2}
	}
	\vspace{-1ex}
	\caption{Example graphs achieving worst-case approximation ratio.}
	\label{fig:two-swap}
	\vspace{-2mm}
\end{figure}

\begin{theorem}\label{thm:lower-bounds}
	For all $k \ge 2$, there is an infinite family of graphs in which the size of a $k$-maximal independent set $I$ is $\frac{2}{\Delta}$ of the optimal.
\end{theorem}

\begin{IEEEproof}
	For $k \in \{2, 3\}$, consider the infinite family of instances given by the complete graphs $K_n$ for $n \ge 4$.
	For each $(u, v) \in K_n$, a vertex $w$ is added between $u$ and $v$, and the edge $(u, v)$ is replaced by two edges $(u, w)$ and $(w, v)$.
	Denote the resulted graph as $K'_n$.
	An example for $k = 3$ and $n = 4$ is shown in Fig.~\ref{fig:worst-case-1}.
	Notice that the original $n$ vertices constitute of a $k$-maximal independent set in $K'_n$.
	However, $\alpha(K'_n) = \binom{n}{2} = \frac{n(n - 1)}{2}$ and $\Delta = n - 1$.
	
	As for $k \ge 4$, consider the infinite family of instances given by the hypercube graphs $Q_n$ for $n \ge k$.
	A hypercube graph $Q_n$ has $2^n$ vertices, and $2^{n-1}n$ edges, and is a regular graph with $n$ edges touching each vertex.
	An illustrating graph $Q_4$ can be found in Fig.~\ref{fig:worst-case-2}.
	%For each edge $(u, v)$ in $Q_n$, a vertex $w$ is added between $u$ and $v$, and the edge $(u, v)$ is replaced by two edges $(u, w)$ and $(w, v)$
	We construct a new graph $Q'_n$ in the same manner as above.
	Since the length of the shortest cycle in $Q_n$ is $n$, the induced graph of any vertex subset $S$ with size $k$ in $Q_n$ has at most $k$ edges.
	Therefore, the original $2^n$ vertices in $Q_n$ constitute of a $k$-maximal independent set in $Q'_n$. 
	However, $\alpha(Q'_n) = 2^{n-1}n$ and $\Delta = n$.
\end{IEEEproof}

Unfortunately, sometimes the above bound may be too loose to use in practice.
Hence, we focus on deriving a more useful lower bound in real-world graphs.
It is observed that the degree distribution of most real-world graphs closely resembles a power law distribution.
And, many graph models capturing this topological property have been proposed for more detailed algorithmic performance analysis~\cite{DBLP:conf/stoc/AielloCL00, DBLP:conf/soda/BrachCLS16, DBLP:journals/algorithmica/ChauhanFR20}.
However, some of them are too strict to describe dynamic graphs, \eg, the power-law random graph model used in~\cite{DBLP:journals/pvldb/LiuLYXW15} which assumes that the number of vertices with degree $d$ is ${e^\alpha}/{d^\beta}$, where $\alpha$ and $\beta$ are two parameters describing the degree distribution.
In what follows, we adopt the power-law bounded graph model proposed in~\cite{DBLP:journals/algorithmica/ChauhanFR20} to make a further analysis of the size of $k$-maximal independent sets.

\begin{definition}[Power-law Bounded Graph Model]\label{def:plb-graph}
	Let $G$ be a $n$-vertex graph and $c_1 > c_2 > 0$ be two universal constants.
	We say that $G$ is power-law bounded (PLB) for some parameters $\beta > 1$ and $t \ge 0$ if for every integer  $\lfloor \log{\delta} \rfloor \le d \le \lfloor \log{\Delta} \rfloor$, where $\delta$ and $\Delta$ denote the minimum and maximum degree in $G$ respectively, the number of vertices $v$ such that $d(v) \in [2^d, 2^{d + 1})$ is at least $c_2n(t+1)^{\beta-1}\sum_{i = 2^d}^{2^{d + 1} - 1}(i + t)^{-\beta}$, and is at most
	$c_1n(t+1)^{\beta-1}\sum_{i = 2^d}^{2^{d + 1} - 1}(i + t)^{-\beta}$.
\end{definition}

The PLB graph model requires that the number of vertices in each buckets $[2^d, 2^{d+1})$ can be bounded by two shifted power-law sequences described by four parameters $c_1, c_2, \beta$, and $t$.
This also holds over dynamic graphs, \ie, the number of vertices with a certain degree may change over time, but the number of vertices with degree in a range can be bounded.
And it is experimentally observed that the majority of real-world networks from the SNAP dataset~\cite{snapnets} satisfy the power-law bounded property with $\beta > 2$~\cite{DBLP:journals/algorithmica/ChauhanFR20}.

\begin{theorem}\label{thm:approx-plb}
	Given a power-law bounded graph $G$ with parameters $\delta = 1$ and $\beta > 2$, if $I$ is an 1-maximal independent set of $G$, then $\alpha(G) \le \min\{\frac{2(t+1)}{c_2}, \frac{2c_1(t + 1)^\beta}{c_2(\beta - 1)(t + 2)^{\beta - 1}} + 1\}|I|$.
\end{theorem}
\begin{IEEEproof}
	Since $I$ is 1-maximal, at least half of the vertices whose degree is one appears in $I$.
	It is derived that
	\begin{displaymath}
		\begin{small}
			2\vert I \vert \ge \vert \{v \in V \mid d(v) = 1\} \vert \ge c_2 n (t + 1)^{-1} \ge \frac{c_2}{t+1} \alpha(G).
		\end{small}
	\end{displaymath}
	Then, it is apparently that the degree of all vertices in $\bar{I}_k$ can not be less than $k$, \ie,
	\begin{displaymath}
		\begin{small}
			\begin{aligned}
			\sum^\Delta_{j = 2} \vert \bar{I}_j \vert & \le \sum^{\lfloor \log{\Delta} \rfloor}_{i = 1} c_1n(t+1)^{\beta - 1}\sum^{2^{i+1}-1}_{j = 2^i}(j + t)^{- \beta}\\
			& \le \frac{2c_1(t + 1)^\beta}{c_2(\beta - 1)(t + 2)^{\beta - 1}} \vert I \vert.
			\end{aligned}
		\end{small}
	\end{displaymath}
	Following the proof of theorem~\ref{thm:oneswap-approx-ratio}, the theorem is proved.
\end{IEEEproof}

The above theorem implies that the size of $I$ is lower-bounded by a parameter-dependent constant multiple of the optimal in most real-world graphs.
%The first bound also holds when $I$ is maximal but not 1-maximal, which is also derived in~\cite{DBLP:journals/algorithmica/ChauhanFR20}.
%However, the second is sometimes better, \eg, in Youtube graph, $\frac{2(t + 1)}{c_2} = 27.6$ and $\frac{2c_1(t + 1)^\beta}{c_2(\beta - 1)(t + 2)^{\beta - 1}} + 1 = 20.8$.

\subsection{$k$-Maximal Independent Set Maintenance}\label{subsec:framework}

\begin{figure*}
	\centering
	\subfigure[Graph before update.]{
		\includegraphics[width=0.19\linewidth]{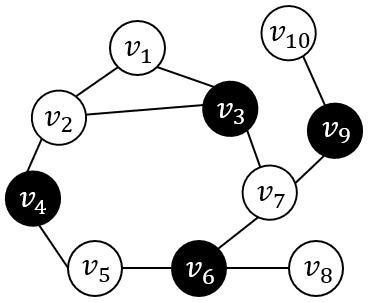}
		\label{fig:origin_graph}
	}\hspace{1ex}
	\subfigure[Information.]{%这个图需要替换掉
		\includegraphics[width=0.24\linewidth]{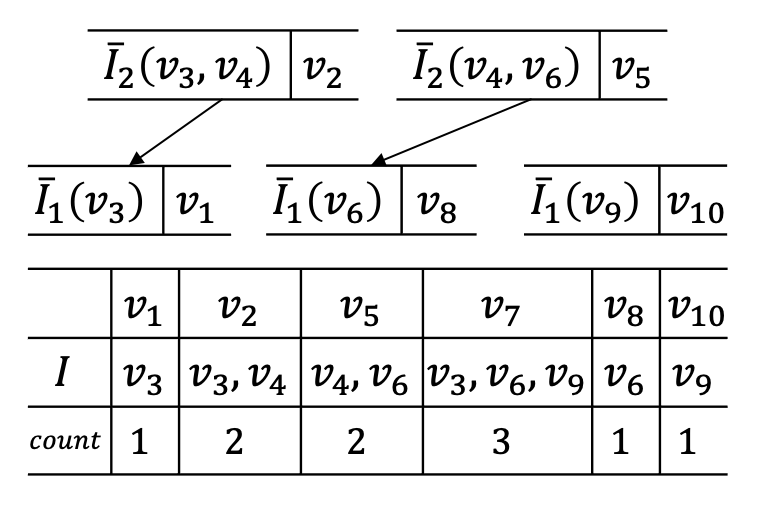}
		\label{fig:info}
	}\hspace{1ex}
	\subfigure[Graph after update ($k = 1$).]{
		\includegraphics[width=0.19\linewidth]{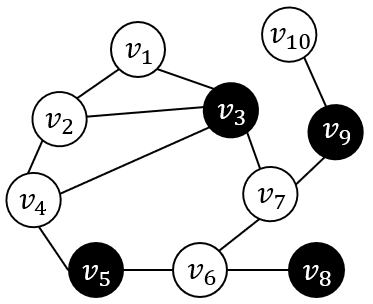}
		\label{fig:alg1}
	}\hspace{1ex}
	\subfigure[Graph after update ($k = 2$).]{
		\includegraphics[width=0.19\linewidth]{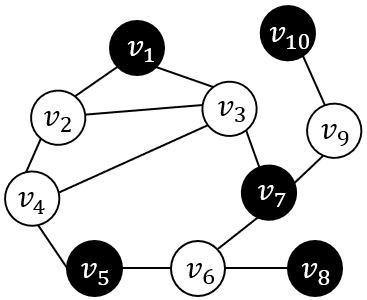}
		\label{fig:alg2}
	}
	\vspace{-1ex}
	\caption{A running example.}
	\label{fig:example}
	\vspace{-4ex}
\end{figure*}

The above analysis indicates that maintaining an 1-maximal independent set addresses the issue of achieving a non-trivial\\ theoretical accuracy guarantee.
Moreover, although considering more kinds of swaps will not make the approximation ratio better, in practice it indeed further improve the quality of the solution, while also increasing the time consumption.
Thus, we introduce a framework that maintains a $k$-maximal independent set for a user-specified $k$.
We start with the information maintained in the framework.

Given a user-specified $k$, let $I$ denote the $k$-maximal independent set maintained by the framework.
Instead of storing $I$ explicitly, the framework keeps a boolean entry $status(v)$ for each vertex $v$ to indicate whether or not it belongs to the current solution.
If required, $I$ will be returned by collecting all vertices whose $status$ is {\sc true}.
And recall that any possible $j$-swap in $I$ for $j \in [k]$ would only appear in the subgraph induced by $I_{\le k} \cup \bar{I}_{\le k}$.
To facilitate the identification of these vertices, for each vertex $v \in \bar{I}$, the framework maintains a list $I(v)$ including the neighbors of $v$ currently in $I$ and a counter $count(v) = |I(v)|$.
And for each subset $S \subseteq I$ of size $j \le k$, it maintains a set $\bar{I}_{\le j}(S) = \{v \in \bar{I}_{\le j} \mid I(v) \subseteq S\}$ of vertices that possibly constitute the swap-in set of $S$.
Since $\bar{I}_{\le j}(S) \subseteq \bar{I}_{\le j'}(S')$ for any two sets $S \subseteq S'$, the framework reorganized all $\bar{I}_{\le j}(S)$s in a hierarchical manner to reduce memory consumption and achieve efficient updates.
That is, for each set $S$ of size $j$, it keeps a list $I_j(S) = \{v \in \bar{I}_j \mid I(v) = S\}$ and pointers to $I_{j-1}(S')$ such that $S' \subset S$.
And if needed, the complete $\bar{I}_{\le j}(S)$ will be collected using a depth-first traversal starting from $I_j(S)$.

Whenever a vertex $v$ is removed from or inserted into $I$, the above information is updated as follows.
The framework iterators over the neighbors $u$ of $v$ to update $I(u)$ accordingly.
After that if $count(u) = j \le k$, it moves $u$ to $\bar{I}_j(I(u))$.
Note that $I(u)$ can be updated in constant time if it is implemented by a doubly-linked list and a pointer to $v \in I(u)$ is recorded in edge $(v, u)$.
And since all $\bar{I}_j(S)$s are disjoint from each other, the {\it hierarchical storage strategy} also allows a constant-time update to the position of $u$ if the index of $u$ in $\bar{I}_j(I(u))$ is maintained explicitly in vertex $u$.
Therefore, the time needed to update the information is bounded by $O(d(v))$.

\begin{example}
	Consider the graph shown in Fig.~\ref{fig:origin_graph}.
	Supposing that the current solution $I = \{v_3, v_4, v_6, v_9\}$, which appears as black vertices, the information maintained in the framework with $k = 2$ is listed in Fig.~\ref{fig:info}.
	According to the hierarchical storage strategy, $v_1$ and $v_8$ is only recorded in $\bar{I}_1(v_3)$ and $\bar{I}_1(v_6)$ respectively.
	If required, $\bar{I}_{\le 2}(v_3, v_4)$ will be collected by merging $\bar{I}_2(v_3, v_4)$ and $\bar{I}_1(v_3)$, and $\bar{I}_{\le 2}(v_4, v_6)$ is returned as $\bar{I}_2(v_4, v_6) \cup \bar{I}_1(v_6)$.
	%The information maintained in the framework with $k = 2$ is listed in the first part of Fig.~\ref{fig:info}.
	%According to the minimal storage strategy, $v_1$ and $v_8$ is only recorded in $\bar{I}_1(v_3)$ and $\bar{I}_1(v_1)$ respectively.
	%And suppose $v_4$ is removed from the solution due to the insertion of edge $(v_3, v_4)$, $(v_6, v_5)$ and $(v_3, v_2)$ will be recorded as candidate vertices in $\cS_1$.
\end{example}

\begin{algorithm}
	\begin{small}
		\caption{\it Framework for Maintenance}
		\label{alg:template}
		\KwIn{A graph $G_{t - 1}$, a $k$-maximal indpendent set $I$ in $G_t$, and an update operation $op$}
		\KwOut{A $k$-maximal independent set $I$ in $G_t$}
		\BlankLine
		$G_t \gets G_{t-1} \oplus op$ and keep $I$ maximal\;
		Collects candidates into $\cS_1, \cdots, \cS_k$ around $op$\;
		\While{$\exists j \in [k]: \cS_j \ne \emptyset$}{
			Let $j$ be the smallest index such that $\cS_j \ne \emptyset$\;
			Retrieve a pair $(S, C(S))$ from $\cS_j$\;
			\ForEach{$v \in C(S)$}{
				\If{$\exists I_S \subseteq \bar{I}_{\le j}(S) \setminus N_t[v] : |I_S| = j$}{
					{\sc MoveOut}($S$); {\sc MoveIn}($\{v\} \cup I_S$)\;
					Extend the solution to be maximal\;
					Find candidates among $\{I(u) \mid u \in N_t[S]\}$\;
				}
			}
			\If{$S$ does not contribute to a $j$-swap {\bf and} $j + 1 \le k$}{
					Find candidates $S' \supset S$ with size $j + 1$\;
			}
		}
		\Return{$\{v \in V_t \mid status(v) = \textsc{true}\}$};
	\end{small}
\end{algorithm}

The details of the framework is presented in Algorithm~\ref{alg:template}.
After updating the structure of the graph, the framework first keeps $I$ to be a maximal independent set in $G_t$ and updates the information accordingly.
Next, the major challenge is to efficiently find all valid swaps caused by the update.
A set $S$ of size $j \le k$ may contribute to a $j$-swap only if some vertices $C(S)$ are newly inserted into $\bar{I}_{\le j}(S)$.
And, the swap-in set $I_S$ must contain at least one vertex in $C(S)$.
The framework collects all such sets $S$ as candidates into $\cS_1, \cdots, \cS_k$ according to their size, respectively.
For each candidate $S \in \cS_j$, a list $C(S)$ including vertices newly added into $\bar{I}_{\le j}(S)$ is also stored in $\cS_j$.
Since now only the information of vertices in the closed neighborhood of $op$ has changed, the framework initializes $\cS_1, \cdots, \cS_k$ among these vertices' neighbors in $I$.
After that, it starts to find swaps in a \textit{bottom-up} manner until all of $\cS_1, \cdots, \cS_k$ are empty.
Concretely speaking, at each loop of the while, let $j \in [k]$ be the smallest integer such that $\cS_j$ is not empty.
The framework retrieves a pair $(S, C(S))$ from $\cS_j$, and for each vertex $v \in C(S)$, it checks whether there exists an independent set $I_S \subseteq \bar{I}_{\le j}(S) \setminus N_t[v]$ of size $j$.
If so, the framework swaps $S$ with $\{v\} \cup I_S$, extends $I$ to be a maximal solution by inserting any vertex in $C(S)$, whose $count$ reduces to zero, into it, and collects new candidates among the closed neighborhood of $S$.
Otherwise, the framework collects new candidates among sets $S' \supset S$ of size $j+1$ into $\cS_{j+1}$ because $C(S) \subseteq \bar{I}_{\le j}(S) \subseteq \bar{I}_{\le j+1}(S')$.
The following theorem guarantees the correctness of the framework.

\begin{theorem}\label{thm:correctness}
	Given a dynamic graph $\mathcal{G}$ and an integer $k$, Algorithm \ref{alg:template} maintains a $k$-maximal independent set $I$ over $\mathcal{G}$.
\end{theorem}

\begin{IEEEproof}
	It is easy to see that an independent set $I$ is $k$-maximal if and only if for each $j \in [k]$, the independence number of the  subgraph induced by every $j$-subsets of $I$ is not greater than $j$.
	Therefore, we prove this theorem by induction that when Algorithm~\ref{alg:template} terminates for $G_t$, $\alpha(G_t[\bar{I}_{\le k}(S)]) \le j$ for every set $S \subseteq I_t$ of size $j \le k$.
	First, we append a $(m_0 + 1)$-length graph sequence $\langle G'_0, \cdots, G'_{m_0} \rangle$ to $\mathcal{G}$, where $G'_0 = (V_0, \emptyset)$, and for each $i \in [m_0]$, $G'_i$ is obtained by inserting an edge in $E_0 \setminus E'_{i-1}$ to $G'_i$.
	This guarantees that there is always a MaxIS $V_0$ in $G'_0$, which is definitely $k$-maximal.
	
	Then, suppose that $I_{t-1}$ is a $k$-maximal independent set in $G_{t-1}$, we claim that $I_t$ is $k$-maximal in $G_t$ when Algorithm~\ref{alg:template} terminates.
	For contradiction, let $S$ be a set that contributes a $j$-swap in $I_t$.
	If $S \subseteq I_{t-1}$, it is known that $\alpha(G_{t-1}[\bar{I}_{\le j}(S)]) \le j$ by the assumption that $I_{t-1}$ is $k$-maximal.
	The increase of $\alpha(G_t[\bar{I}_{\le j}(S)])$ in $G_t$ is because some vertices are newly added to $\bar{I}_{\le j}(S)$.
	Otherwise, $\bar{I}_{\le j}(S)$ is empty in $G_{t-1}$ but not during the update.
	In both of these two cases, $S$ would be inserted into $\cS_j$, which contradicts to the terminal condition of Algorithm~\ref{alg:template}.
	This completes the proof.
\end{IEEEproof}

The preceding analysis implies that the maintained result is a $(\frac{\Delta}{2} + 1)$-approximate MaxIS, and even a constant approximation of the MaxIS if $\mathcal{G}$ is power-law bounded.

\begin{theorem}
	Given a dynamic graph $\mathcal{G}$ and an integer $k$, Algorithm~\ref{alg:template} maintains a $(\frac{\Delta}{2} + 1)$-approximate maximum independent set over $\mathcal{G}$.
	Moreover, if $\mathcal{G}$ is a power-law bounded dynamic graph with $\delta = 1$ and $\beta > 2$, the approximation ratio achieved by algorithm~\ref{alg:template} is $\min\{\frac{2(t + 1)}{c_2}, \frac{2c_1(t + 1)^\beta}{c_2(\beta - 1)(t + 2)^{\beta - 1}} + 1\}$, which is a parameter-dependent constant.
\end{theorem}%\vspace{1ex}

\noindent{\bf Discussions.} We discuss the novelty of the framework, and some strategies that can be used to improve the performance.

\noindent\underline{\it Novelty of the Framework.}
Despite swap operations have been used to improve the quality of a resultant independent set on static graphs in~\cite{DBLP:journals/heuristics/AndradeRW12, DBLP:journals/pvldb/LiuLYXW15}, the framework is superior for the following two reasons.
Firstly, following the framework, it is easy to instantiate an algorithm that efficiently maintains a $k$-maximal independent set over dynamic graphs.
The information maintained by the framework and the bottom-up searching procedure ensures the efficiency and effectiveness of finding all valid swaps after each update.
And the hierarchical storage strategy enables efficient update of the information under dynamic setting while reducing memory consumption.
Second, the bottom-up searching manner guarantees that the current solution $I$ is $(j-1)$-maximal when handling a candidate $S$ of size $j$.
Therefore, some useful properties can be derived to further reduce the search space for $I_S$, \eg, when instantiating the algorithm for $k = 2$, a subset of $C(S)$ will be checked without missing any 2-swaps.

\noindent\underline{\it Optimization Techniques.}
Two strategies are found that can be used to further improve the performance of the framework.
1) Recall that the framework maintains a list $I(v)$ for each vertex $v \in \bar{I}$ including all its neighbors currently in $I$.
However, it is noticed that only the list $I(v)$ of vertices $v$ with $count(v) \le k$ will be actually used during the update procedure.
A {\it lazy collection strategy} could benefit a lot in the scenario of small $k$.
That is, the framework only maintains $count$ for each vertex explicitly, and collects other information in real time if needed.
But the worst-case time complexity of an algorithm with such strategy can not be well bounded.
2) {\it Perturbation} is a classical method to help local search methods get rid of a local optima.
Many random strategies are proposed to find a better solution~\cite{DBLP:books/sp/03/LourencoMS03}.
However, it is important to balance the effectiveness and the time consumption under dynamic setting.
With the intuition that high-degree vertices are less likely to appear in a MaxIS, a solution vertex may be swapped with its smallest-degree neighbor in $\bar{I}_1$ while finding valid swaps.
	\section{Two Dynamic Algorithms}\label{sec:oneswap}

In this section, we instantiate two dynamic algorithms by setting $k = 1$ and $k = 2$ in the framework respectively.

\subsection{Dynamic OneSwap Algorithm}

Following the framework, we propose an algorithm that maintains an 1-maximal independent set over dynamic graphs.
Recall that an 1-swap consists of removing one vertex from the solution $I$ and inserting at least two vertices into it.
Since any possible 1-swap can only appear in $G[I_1 \cup \bar{I}_1]$, where $\bar{I}_1 = \{v \in \bar{I} \mid count(v) = |N(v) \cap I| = 1\}$ and $I_1 = \{v \in I \mid \\ N(v) \cap \bar{I}_1 \ne \emptyset\}$, the algorithm maintains a list $\bar{I}_1(v) = \{u \in N(v) \mid count(u) = 1\}$ for each vertex $v \in I$.
It is apparently that $I$ is an 1-maximal independent set if and only if $G[\bar{I}_1(v)]$ is a clique for all vertices $v \in I$.
And, a vertex $v \in I$ may contribute to an 1-swap only if some vertices $C(v)$ are newly added to $\bar{I}_1(v)$.
We call such vertices candidates in the following.
The algorithm uses $\cS_1$ to store all candidates $v$ along with their $C(v)$ during the update procedure.

The pseudocode is shown in Algorithm~\ref{alg:oneswap}.
Assuming that an update $op$ is performed on $G_{t-1}$, the algorithm first keeps the solution $I$ maximal and collects candidates to $\cS_1$ as follows.
\begin{enumerate}[$\bullet$]
\item In case of inserting a vertex $v$, it iterates over $N_t(v)$ to compute $count(v)$ and inserts $v$ into $I$ if $count(v) = 0$, or inserts $v$ into $C(I(v))$ and $I(v)$ into $\cS_1$ if $count(v) = 1$.
\item In case of deleting a vertex $v \in I$, it removes $v$ from $I$, and inserts any neighbor of $v$, whose $count$ reduces to zero, into $I$.
Then, for each vertex $u \in N_{t-1}(v)$ with $count(u) = 1$, it inserts $u$ into $C(I(u))$ and $I(u)$ into $\cS_1$.
\item On insertion of an edge $(u, v)$ between two vertices in $I$, if one of them, say $v$, with $\bar{I}_1(v) \ne \emptyset$, it removes $v$ from $I$, and inserts any neighbor of $v$, whose $count$ reduces to zero, into $I$.
Otherwise, it removes the vertex with higher degree, say $v$, from $I$.
%It inserts $v$ into $C(u)$ and $u$ into $\cS_1$ if $count(v) = 1$.
Then, for each $w \in N_{t-1}[v]$ with $count(w) = 1$, it inserts $w$ into $C(I(w))$ and $I(w)$ into $\cS_1$.
\item On deletion of an edge $(u, v)$, there are two cases to consi-der.
i) If one of them, say $u$, belongs to $I$, it removes $u$ from $I(v)$ and inserts $v$ into $I$ if $count(v) = 0$, or inserts $v$ into $C(I(v))$ and $I(v)$ into $\cS_1$ if $count(v) = 1$.
ii) If neither $u$ nor $v$ is in $I$ and $I(u) = I(v) = w$, it removes $w$ from $I$ and inserts $u, v$ into $I$.
Then, for each $x \in N_t(w)$ with $count(x) = 1$, it inserts $x$ into $C(I(x))$ and $I(x)$ into $\cS_1$. 
\end{enumerate}
After that, the algorithm checks whether $G_t[\bar{I}_1(v)]$ is still a clique for each candidate $v$ recorded in $\cS_1$.
Concretely speaking, for each vertex $u \in C(v)$, it calculates the number of $u$'s closed neighbors appears in $\bar{I}_1(v)$.
If $\vert N_t[u] \cap \bar{I}_1(v) \vert < \vert \bar{I}_1(v) \vert$,  then $G_t[\bar{I}_1(v)]$ is no more a clique.
The algorithm removes $v$ from $I$ and inserts $u$ into it.
Next it extends $I$ to be maximal by inserting any vertex in $\bar{I}_1(v)$, whose $count$ reduces to zero, into $I$.
Finally, for each vertex in $N_t(v)$ whose $count$ reduces to one, the algorithm marks its neighbor in $I$ as new candidates.
And the algorithm terminates when $\cS_1$ is empty.

\begin{algorithm}[htb]
	\begin{small}
		\caption{\it Dynamic OneSwap Algorithm}
		\label{alg:oneswap}
		\KwIn{A graph $G_{t - 1}$, an 1-maximal independent set $I$ in $G_{t - 1}$, and an update operation $op$}
		\KwOut{An 1-maximal independent set $I$ in $G_t$}
		\BlankLine
		$G_t \gets G_{t-1} \oplus op$ and keep the solution $I$ maximal\;
		Collect candidates into $\cS_1$ around $op$\;
		\While{$\cS_1 \ne \emptyset$}{
			Retrieve a pair $(v, C(v))$ from $\cS_1$\;
			\ForEach{$u \in C(v)$}{
				\If{$\vert N_t[u] \cap \bar{I}_1(v) \vert < \vert \bar{I}_1(v) \vert$}{
					{\sc MoveOut}($v$); {\sc MoveIn}($u$)\;
					\ForEach{$w \in \bar{I}_1(v)$}{
						\lIf{$count(w) = 0$}{ {\sc MoveIn}($w$)}
					}
					\ForEach{$w \in N_t[v]$ {\bf s.t.} $count(w) = 1$}{
						Insert $w$ into $C(I(w))$ and $I(w)$ into $\cS_1$\;
					}
				}
			}
		}
		\Return{$\{v \in V_t \mid status(v) = \textsc{true}\}$};
	\end{small}
\end{algorithm}

\begin{example}\label{eg:oneswap}
	Consider the graph shown in Fig.~\ref{fig:origin_graph} and the information shown in Fig.~\ref{fig:info}.
	After the edge $(v_3, v_4)$ is inserted, Algorithm~\ref{alg:oneswap} first removes $v_4$ from $I$.
	Then, since both $count(v_2)$ and $count(v_5)$ reduce to one, it collects $v_6$ and $v_3$ as candidates into $\cS_1$ with $C(v_6) = \{v_5\}$ and $C(v_3) = \{v_2, v_4\}$.
	Because $\vert N[v_5] \cap \bar{I}_1(v_6) \vert = 1 < \vert \bar{I}_1(v_6) \vert = 2$, the algorithm swaps $v_6$ with $v_5$ and extends $I$ to be maximal by inserting $v_8$ into it.
	After that, the algorithm stops since $\cS_1$ is empty, and the final result is shown in Fig.~\ref{fig:alg1}.
\end{example}%\smallskip

\noindent{\bf Performance Analysis.}
At each loop, Algorithm~\ref{alg:oneswap} retrieves a pair $(v, C(v))$ from $\cS_1$ and calculates $\vert N_t[u] \cap \bar{I}_1(v) \vert$ for each vertex $u \in C(v)$ to determine whether or not $G_t[\bar{I}_1(v)]$ is still a clique.
This can be accomplished in $O(\sum_{u \in C(v)} d_t(u))$ time because $I(v)$ is maintained explicitly for each vertex $v \in \bar{I}$.
Therefore, if $v$ does not contribute to an 1-swap, the time consumption is at most $O(\sum_{u \in C(v)} d_t(u))$.
Otherwise, let $u$ be the vertex in $C(v)$ such that $\vert N_t[u] \cap \bar{I}_1(v) \vert < \vert \bar{I}_1(v) \vert$, the algorithm takes $O(d_t(v) + d_t(u))$ time to swap $v$ and $u$, extends $I$ to be a maximal solution in at most $O(\sum_{w \in \bar{I}_1(v)}d_t(w))$ time and collects new candidates in $O(d_t(v))$ time.
Since the newly added vertex sets of any two candidates are disjoint, the time complexity of Algorithm~\ref{alg:oneswap} is $O(\sum_{v \in I}{(d_t(v) + \sum_{u \in \bar{I}_1(v)}{d_t(u)})}) = O(\sum_{v \in I \cup \bar{I}_1} d_t(v)) = O(m_t)$.
And according to Theorem~\ref{thm:correctness}, Algorithm~\ref{alg:oneswap} maintains an 1-maximal independent set over dynamic graphs.
Thus the approximation ratio of Algorithm~\ref{alg:oneswap} is $\frac{\Delta_t}{2} + 1$, where $\Delta_t$ is the maximum vertex degree of the current graph $G_t$.

As for power-law bounded graphs with parameters $\delta = 1$ and $\beta > 2$,  the approximation ratio achieved by Algorithm~\ref{alg:oneswap} is $\min\{\frac{2(t + 1)}{c_2}, \frac{2c_1(t + 1)^\beta}{c_2(\beta - 1)(t + 2)^{\beta - 1}} + 1\}$, which is a constant depending on the parameters.
And with the Lemma 3.5 of~\cite{DBLP:conf/soda/BrachCLS16}, we know that the time complexity of Algorithm~\ref{alg:oneswap} is $O((1 + t)n_t)$.

	\subsection{Dynamic TwoSwap Algorithm} \label{sec:twoswap}

Although it is proved that considering more kinds of swaps will not improve the approximation ratio, finding 2-swaps in an independent set can indeed further enlarge its size in the absence of 1-swap.
Hence, we instantiate an algorithm that maintains a 2-maximal independent set.

Given a graph $G$ and an independent set $I$ in $G$, a 2-swap consists of removing two vertices from $I$ and inserting at least three vertices into it.
Any possible 2-swap can only appear in $G[I_{\le 2} \cup \bar{I}_{\le 2}]$, where $\bar{I}_{\le 2} = \{v \in \bar{I} \mid count(v) \le 2\}$ and $I_{\le 2} = \{v \in I \mid N(v) \cap \bar{I}_{\le 2} \ne \emptyset\}$.
Following the framework,\\ the algorithm maintains all vertices in $\bar{I}_{\le 2}$ explicitly in a hierarchical structure, and finds 2-swaps when the maintained solution $I$ is 1-maximal.
This suggests that a set $S$ contribute to a 2-swap if and only if there is an independent set $I_S \subseteq\\ I_{\le 2}(S)$ of size three, which must contains a vertex $x \in \bar{I}_2(S)$.
Therefore, the algorithm only records vertices with $count = 2$ in all $C(S)$ to further reduce the search space.

The pseudocode is shown in Algorithm~\ref{alg:twoswap}.
After performing the update $op$ on $G_{t-1}$, the algorithm updates $I$ as a maximal solution and collects candidates to $\cS_1$ and $\cS_2$ as follows.
\begin{enumerate}[$\bullet$]
	\item In case of inserting a vertex $v$, it iterates over $N_t(v)$ to compute $count(v)$ and inserts $v$ into $I$ if $count(v) = 0$, or inserts $v$ into $C(I(v))$ and $I(v)$ into $\cS_i$ if $count(v) = i \le 2$.
	\item In case of deleting a vertex $v \in I$, it removes $v$ from $I$, and inserts any neighbor of $v$, whose $count$ reduces to zero, into $I$.
	Then, for each vertex $u \in N_{t-1}(v)$ with $count(u) = i \le 2$, it inserts $u$ into $C(I(u))$ and $I(u)$ into $\cS_i$.
	\item On insertion of an edge $(u, v)$ between two vertices in $I$, if one of them, say $v$, with $\bar{I}_1(v) \ne \emptyset$, it removes $v$ from $I$, and inserts any neighbor of $v$, whose $count$ is zero, into $I$.
	Otherwise, it removes the one with higher degree, say $v$, from $I$.
	%It inserts $v$ into $C(I(v))$ and $I(v)$ into $\cS_i$ if $count(v) = i \le 2$.
	Then, for each $w \in N_{t-1}[v]$ with $count(w) = i \le 2$, it inserts $w$ to $C(I(w))$ and $I(w)$ into $\cS_i$.
	\item On deletion of an edge $(u, v)$, there are two cases to consider.
	i) if one of them, say $u$, belongs to $I$, it removes $u$ from $I(v)$ and inserts $v$ into $I$ if $count(v) = 0$, or inserts $v$ into $C(I(v))$ and $I(v)$ into $\cS_i$ if $count(v) = i\le 2$.
	ii) if neither $u$ nor $v$ is in $I$, the following three cases are considered.
	a) If $I(u) = I(v) = w$, it swaps $w$ with $u, v$.
	Then for each $x \in N_t(w)$ with $count(x) = i \le 2$, it inserts $x$ into $C(I(x))$ and $I(x)$ into $\cS_i$.
	b) If $I(u) = x \ne I(v) = y$ and there exists a vertex $w \in \bar{I}_2(x, y)$ such that edges $(u, w)$ and $(v, w)$ are not in $E_t$, it swaps $x, y$ with $u, v, w$.
	Then for each $z \in N_t(x, y) $ with $count(z) = i \le 2$, it inserts $z$ into $C(I(z))$ and $I(z)$ into $\cS_i$.
	c) If $I(u) \subseteq I(v) = \{x, y\}$, it inserts $v$ into $C(I(v))$ and $I(v)$ into $\cS_2$.
\end{enumerate}
After that, Algorithm~\ref{alg:twoswap} finds 1-swaps as stated in Algorithm~\ref{alg:oneswap} if $\cS_1$ is not empty.
Additionally, if a candidate $v$ does not contribute to an 1-swap, it collects new candidates, which are superset of $v$, to $\cS_2$ as shown in line 14 -17.
When $\cS_1$ is empty but $\cS_2$ is not, the algorithm retrieves a pair $(S, C(S))$ from $\cS_2$, say $S = \{u, v\}$.
For each vertex $x \in C(S)$, it checks if there exists a triangle $(x, y, z)$ in the complement of $G_t[\bar{I}_{\le 2}(S)]$.
Since the solution $I$ is now 1-maximal and $y, z$ are not adjacent to $x$, the algorithm refines the candidate vertex sets of $y, z$ to $C_y = \bar{I}_1(u) \cup \bar{I}_2(S) \setminus N_t[x]$ and $C_z = \bar{I}_1(v) \cup \bar{I}_2(S) \setminus N_t[x]$, respectively.
Then for each vertex $y \in C_y$, it calculates the number of $y$'s closed neighbors appears in $C_z$.
If $\vert N_t[y] \cap C_z \vert < \vert C_z \vert$, the algorithm removes $u, v$ from $I$ and inserts $x, y$ into it, and extends $I$ to be a maximal solution by inserting any vertex in $\bar{I}_{\le 2}(S)$, whose $count$ reduces to zero, into $I$.
Finally, for each vertex in $N_t[S]$ whose $count$ reduces to two or less, the algorithm marks its neighbor(s) in $I$ as new candidates.
And the algorithm terminates when both $\cS_1$ and $\cS_2$ are empty.

\begin{algorithm}[htb]
	\begin{small}
		\SetKwProg{proc}{Procedure}{}{}
		\caption{\it \small Dynamic  TwoSwap Algorithm}
		\label{alg:twoswap}
		\KwIn{The graph $G_{t-1}$, a 2-maximal independent set $I$ in $G_{t-1}$, and an update operation $op$}
		\KwOut{A 2-maxiaml independent set $I$ of $G_t$}
		\BlankLine
		$G_t \gets G_{t-1} \oplus op$ and keep the solution $I$ maximal\;
		Collect candidates into $\cS_1$ and $\cS_2$ around $op$\;
		\While{$\cS_1 \ne \emptyset$ {\bf or} $\cS_2 \ne \emptyset$}{
			\lIf{$\cS_1 \ne \emptyset$}{{\sc FindOneSwap}()}
			\lElseIf{$\cS_2 \ne \emptyset$}{{\sc FindTwoSwap}()}
		}
		\Return{$\{v \in V_t \mid status(v) = \textsc{true}\}$};
		\BlankLine
		\proc{{\sc FindOneSwap}\emph{()}}{
			Retrieve a pair $(v,C(v))$ from $\cS_1$\;
			\If{$\exists u \in C(v) : \vert N_t[u] \cap \bar{I}_1(v) \vert < \vert \bar{I}_1(v) \vert$}{
				{\sc MoveOut}($v$); {\sc MoveIn}($u$)\;
				\ForEach{$w \in \bar{I}_1(v)$}{
					\lIf{$count(w) = 0$}{{\sc MoveIn}($w$)}
				}
				{\sc FindCandidates}($v$)\;	
			}
			\Else{
				\ForEach{$u \in \bar{I}_{2}(v)$}{
					\If{$\vert N_t(u) \cap C(v) \vert < \vert C(v) \vert$}{Insert $u$ into $\cS_2(I(u))$\;}
				}
			}
		}
		\BlankLine
		\proc{{\sc FindTwoSwap}\emph{()}}{
			Retrieve a pair $(S, C(S))$ from $\cS_2$, and let $S = \{u, v\}$\;
			\ForEach{$x \in C(S)$}{
				$C_y \gets \bar{I}_1(u) \cup \bar{I}_2(S) - N_t[x]$\;
				$C_z \gets \bar{I}_1(v) \cup \bar{I}_2(S) - N_t[x]$\;
				\ForEach{$y \in C_y$}{
					\If{$\vert N_t[y] \cap C_z \vert < \vert C_z \vert$}{
						{\sc MoveOut}($S$); {\sc MoveIn}($x, y$)\;
						\ForEach{$w \in \bar{I}_{\le 2}(S)$}{
							\lIf{$count(w) = 0$}{{\sc MoveIn}($w$)}	
						}
						{\sc FindCandidates}($S$)\;
					}
				}
			}
		}
		\BlankLine
		\proc{{\sc FindCandidates}\emph{($S$)}}{
			\ForEach{$u \in N_t[S]$ {\bf s.t.} $count(u) = i \le 2$}{
				Insert $u$ into $C(I(u))$ and $I(u)$ into $\cS_i$\;
			}	
		}
	\end{small}
\end{algorithm}

\begin{example}
	Continue with Example~\ref{eg:oneswap} with $k = 2$.
	After sw-apping $v_6$ with $v_5, v_8$, Algorithm~\ref{alg:twoswap} collects $\{v_3, v_5\}$, $\{v_5, v_8\}$, and $\{v_3, v_9\}$ as new candidates into $\cS_2$ with $C(\{v_3, v_5\}) = \{v_4\}$, $C(\{v_5, v_8\}) = \{v_6\}$, and $C(\{v_3, v_9\}) =  \{v_7\}$.
	Since $\cS_1$ is empty, the algorithm retrieves a pair, say $(\{v_3, v_9\}, v_7)$, from $\cS_2$.
	Then it computes $C_y = \{v_1, v_2\}$ and $C_z = \{v_{10}\}$, and finds $\vert N[v_1] \cap C_z\vert = 0 < \vert C_z \vert = 1$.
	The algorithm swaps $v_3, v_9$ with $v_1, v_7$ and extends $I$ to be maximal by inserting $v_{10}$ into it.
	After that, since both $\cS_1$ and $\cS_2$ are empty, the algorithm stops and the final result is shown in Fig.~\ref{fig:alg2}.
\end{example}

\noindent{\bf Performance Analysis.}
According to Theorem~\ref{thm:correctness}, Algorithm~\ref{alg:twoswap} always maintains a 2-maximal independent set over dynamic graphs.
Thus the approximation ratio of Algorithm~\ref{alg:twoswap} is $\frac{\Delta_t}{2} + 1$, where $\Delta_t$ is the maximum vertex degree in $G_t$.
As for power-law bounded graphs with parameters $\delta = 1$ and $\beta > 2$,  the approximation ratio is $\min\{\frac{2(t + 1)}{c_2}, \frac{2c_1(t + 1)^\beta}{c_2(\beta - 1)(t + 2)^{\beta - 1}} + 1\}$, which is a parameter-dependent constant.

Since candidates recorded in $\cS_1$ are handled in the same way as stated in Algorithm~\ref{alg:oneswap}, we focus on the time consumed by all candidates collected in $\cS_2$.
When the solution $I$ is 1-maximal, Algorithm~\ref{alg:twoswap} retrieves a pair $(S, C(S))$ from $\cS_2$.
For each vertex $x \in C(S)$, it first takes $O(|\bar{I}_{\le 2}(S)| + d_t(x))$ time to build the candidate sets $C_y$ and $C_z$.
Then it calculates $\vert N_t[y] \cap C_z \vert < \vert C_z \vert$ in $O(d_t(y))$ time for each vertex $y \in C_y$.
If $S$ does not contribute to a 2-swap, the time consumption is at most $O(\sum_{x \in C(S)}(d_t(x) + \vert \bar{I}_{\le 2}(S) \vert +\sum_{y \in C_y} d_t(y)))$.
Otherwise, let $x$ be a vertex in $C(S)$ with two non-adjacent vertices $y \in C_y$ and $z \in C_z$.
Algorithm~\ref{alg:twoswap} takes $O(d_t(S))$ time to remove $S$ form $I$ and $O(d_t(x) + d_t(y))$ time to insert $x, y$ into it.
After that, it takes at most $O(\sum_{z \in \bar{I}_{\le 2}(S)} d_t(z))$ time to extend $I$ to be maximal and at most $O(d_t(S))$ time to collect new candidates to $\cS_1$ and $\cS_2$.
Since the $count$ of each vertex collected in $C(S)$ is two, the $C(S)$ of any two candidates $S$ in $\cS_2$ are disjoint.
Therefore, the time complexity of Algorithm~\ref{alg:twoswap} can be bounded by $O(\tau\sum_{v \in I \cup \bar{I}_{\le 2}}d_t(v)) = O(\tau m_t)$, where $\tau = \max_{v \in I}{\vert \bar{I}_2(v) \vert}$.

We also make a further analysis of the expected value of $|\bar{I}_2(v)|$ for a vertex $v$ in $I$ on a power-law bounded graph.
The randomness comes from the generation of edges in the graph.
We adopt the erased configuration model here, which is widely used for generating a random network from a given degree sequence.
Specifically, the model generates $d(v)$ stubs for each vertex $v$, and then matches them independently uniformly at random to create edges.
Finally, loops and multiple edges are removed in order to generate a simple graph.

\begin{lemma}
Given a power-law bounded graph $G$,
\begin{equation*}
	\begin{small}
		{\rm E}[|\bar{I}_2(v)|] \le \frac{c_1(t+1)^\beta}{2c_2}(\zeta(2\beta - 4)\bar{d})^\frac{1}{2},
	\end{small}
\end{equation*}
where $\zeta(\beta)$ is the Riemann zeta function with parameter $\beta$, and $\bar{d}$ is the average degree of $G$.
\end{lemma}

\begin{IEEEproof}
	Given a power-law bounded graph $G = (V, E)$, let $V_i = \{v \in V \mid d(v) = i\}$ and $n_i = |V_i|$.
	Suppose that $I$ is a maximal independent set in $G$, and define $\psi = \sum_{v \in I} d(v)$.
	It is easy to derive that $\frac{1}{2} n \le \psi \le \frac{\bar{d}}{2}n$, where $\bar{d}$ is the average degree of $G$.
	For a vertex $v \in I$ with degree $d$, define a sequence of boolean random variables $X_1, \cdots, X_d$, where $X_i = 1$ if and only if the $i$-th stub of $v$ is matched with a stub of a vertex whose $count$ is two.
	Thus, we have
	\begin{displaymath}
		\begin{small}
			\begin{aligned}
				\Pr\{X_i = 1 \mid d(v) = d\} \le \sum^\Delta_{i = 2} \frac{i\cdot n_i}{2m} \frac{(i - 1)(\psi - d)}{2m} (\frac{2m - 2\psi}{2m})^{i - 2}
			\end{aligned}
		\end{small}
	\end{displaymath}
	Then, with the law of total expectation, we have
	\begin{displaymath}
		\begin{small}
			\begin{aligned}
				{\rm E}[|\bar{I}_2(v)|] =& \sum^\Delta_{d = 1} {\rm E}[|\bar{I}_2 \cap N(v)| \mid d(v) = d] \cdot \Pr\{d(v) = d\} \\
				\le& \sum^\Delta_{d = 1} \frac{|I \cap V_d|}{|I|}d\cdot \sum^\Delta_{i = 2} \frac{i\cdot n_i}{2m} \frac{(i - 1)(\psi - d)}{2m} (\frac{2m - 2\psi}{2m})^{i - 2}\\
				\le& \frac{\psi^2}{4m^2|I|}\sum^\Delta_{i = 2}n_i \times i^2(1 - \frac{\psi}{m})^{i - 2}
			\end{aligned}
		\end{small}
	\end{displaymath}
	According to Lemma 3.3 in~\cite{DBLP:conf/soda/BrachCLS16}, and it is easy to see that $f(ci) = (ci)^2(1 - \frac{\psi}{m})^{ci - 2} \le c^2f(i)$, thus
	\begin{displaymath}
		\begin{small}
			\begin{aligned}
				{\rm E}[|\bar{I}_2(v)|] \le &\frac{\psi^2}{4m^2|I|}\times  4c_1 n(t+1)^{\beta - 1}\sum_{i = 2}^\Delta(i + t)^{-\beta}i^2(1 - \frac{\psi}{m})^{i - 2}\\
				\le & \frac{c_1(t + 1)^\beta}{2c_2}\sum_{i = 2}^\Delta(i + t)^{-\beta}i^2(1 - \frac{\psi}{m})^{i - 2}\\
				\le & \frac{2c_1(t + 1)^\beta}{2c_2} (\zeta(2\beta - 4)\bar{d})^\frac{1}{2}
			\end{aligned}
		\end{small}
	\end{displaymath}
	The second inequality is because $2|I| \ge c_2n(t + 1)^{-1}n$ as stated in the proof of theorem~\ref{thm:oneswap-approx-ratio}, and the last inequality is due to the Cauchy-Schwarz inequality.
\end{IEEEproof}

And, with the fact that $m_t = O((t+1)n_t)$, we conclude that the expected time complexity of Algorithm~\ref{alg:twoswap} on a power-law bounded graph with parameter $\delta = 1$ and $\beta > 2$ is $O(c_1c_2^{-1}(t+1)^{\beta + \frac{1}{2}}\zeta(2\beta - 4)^\frac{1}{2}n_t)$.

	\section{Experiments}\label{sec:exp}

In this section, we conduct extensive experiments to evaluate the efficiency and effectiveness of the proposed algorithms.

\subsection{Experiment Setting}

\noindent{\bf Datasets.}
22 real graphs are used to evaluate the algorithms.
All of these graphs are downloaded form the Stanford Network Analysis Platform\footnote{http://snap.stanford.edu/data/}~\cite{snapnets} and Laboratory for Web Algorithmics\footnote{http://law.di.unimi.it/datasets.php}~\cite{BoVWFI, BRSLLP}. 
The statistics  are summarized in Table \ref{tab:graphs}, where the last column gives the average degree $\bar{d}$ of each graph.
The graphs are categorized into easy instances and hard instances according to whether a MaxIS in it can be returned by {\small \sf VCSolver}~\cite{DBLP:journals/tcs/AkibaI16} within five hours, and the easy instances are listed in the first half of Table \ref{tab:graphs}.

\begin{table}[htb]\vspace{-1ex}
	\centering
	\caption{Statistics of graphs}\vspace{-1ex}
	\label{tab:graphs}
	\begin{tabular}{c|c|c|c}
		\toprule
		Graph & $n$ & $m$ & $\bar{d}$\\
		\hline
		Epinions	&	75,879 	&	405,740	&	10.69	\cr
		Slashdot	&	82,168 	&	504,230	&	12.27 	\cr
		Email & 265,214 & 364,481 & 2.75 \cr
		com-dblp	&	317,080 	&	1,049,866	&	6.62	\cr
		com-amazon	&	334,863 	&	925,872	&	5.53	\cr
		web-Google	&	875,713 	&	4,322,051	&	9.87	\cr
		web-BerkStan	&	685,230 	&	6,649,470	&	19.41 \cr
		in-2004 & 1,382,870 & 13,591,473 & 19.66 \cr
		as-skitter	&	1,696,415 	&	11,095,298	&	13.08 \cr
		hollywood & 1,985,306 & 114,492,816 & 115.34 \cr
		WikiTalk	&	2,394,385 	&	4,659,565	&	3.89 	\cr
		com-lj	&	3,997,962 	&	34,681,189	&	17.35	\cr
		soc-LiveJournal	&	4,847,571 	&	42,851,237	&	17.68 	\cr
		\hline
		soc-pokec	&	1,632,803 	&	22,301,964	&	27.32 	\cr
		wiki-topcats	&	1,791,489 	&	25,444,207	&	28.41 \cr
		com-orkut	&	3,072,441 	&	117,185,083	&	76.28 \cr
		cit-Patents	&	3,774,768 	&	16,518,947	&	8.75 \cr
		uk-2005 & 39,454,746 & 783,027,125 & 39.70 \cr
		it-2004 & 41,290,682 & 1,027,474,947 & 49.77 \cr
		twitter-2010	&	41,652,230 	&	1,468,365,182	&	70.51 	\cr
		Friendster	&	65,608,366 	&	1,806,067,135	&	55.06	\cr
		uk-2007 & 109,499,800 & 3,448,528,200 &  62.99 \cr
		\bottomrule
	\end{tabular}
	\vspace{-2mm}
\end{table}

\noindent{\bf Algorithms.}
We implement the following two algorithms,
\begin{enumerate}[$\bullet$]
	\item \textit{\small DyOneSwap}: the dynamic $(\frac{\Delta}{2} + 1)$-approximation algorithm that maintains an 1-maximal independent set,
	\item \textit{\small DyTwoSwap}: the dynamic $(\frac{\Delta}{2} + 1)$-approximation algorithm that maintains a 2-maximal independent set,
\end{enumerate}
and compare them with the state-of-the-art methods {\small\it DGOneDIS} and {\small\it DGTwoDIS} proposed in~\cite{DBLP:conf/icde/ZhengPCY19}, which maintain a near-maximum independent set over dynamic graphs without theoretical accuracy guarantees, and the dynamic version {\small\it DyARW} of {\small \sf ARW} proposed in~\cite{DBLP:journals/heuristics/AndradeRW12}, which also uses 1-swaps to improve the size of independent sets on static graphs.
All the algorithms are implemented in C++ and complied by GNU G++ 7.5.0 with -O2 optimization;
the source codes of \textit{\small DGOneDIS} and \textit{\small DGTwoDIS} are obtained from the authors of~\cite{DBLP:conf/icde/ZhengPCY19} while all other algorithms are implemented by us.
All experiments are conducted on a machine with a 3.5-GHz Intel(R) Core(TM) i9-10920X CPU, 256GB main memory, and 1TB hard disk running CentOS 7.
Similar to~\cite{DBLP:conf/icde/ZhengPCY19}, we randomly insert/remove a predetermined number of vertices/edges to simulate the update operations.
For easy graphs, we uses a MaxIS computed by {\small \sf VCSolver}~\cite{DBLP:journals/tcs/AkibaI16} as the initial independent set, and for hard graphs, we treat the independent set returned by {\small \sf ARW}~\cite{DBLP:journals/heuristics/AndradeRW12} as the input one.
This is reasonable since all initial independent sets are obtained within limited time consumption.	

\noindent{\bf Metrics.}
We evaluate all these algorithms from the following three aspects: size of the maintained independent set, response time, and memory usage.
Firstly, the larger the size of the independent set maintained by an algorithm, the better the algorithm; we report the gap and the accuracy achieved by each algorithm in our experiments.
Secondly, for the response time, the smaller the better; we run each algorithm three times and report the average CPU time.
Thirdly, the smaller memory consumed by an algorithm the better; we measure the heap memory usage by the command {\small \sf/usr/bin/time}\footnote{https://man7.org/linux/man-pages/man1/time.1.html}.

\subsection{Experimental Results}

We report our findings concerning the performance of these algorithms in this section.

\begin{table*}[htb]
	\centering
	\caption{The gap to the independence number obtained by {\sf VCSolver}~\cite{DBLP:journals/tcs/AkibaI16} and accuracy on easy graphs after 100,000 updates.}
	\vspace{-2ex}
	\label{tab:easy-gap}
	\begin{tabular}{c|c|cc|cc|cc|ccc|ccc}
		\toprule
		\multirow{2}{*}{Graphs} & \multirow{2}{*}{$\alpha(G)$}& \multicolumn{2}{c|}{\textit{DGOneDIS}} & \multicolumn{2}{c|}{\textit{DGTwoDIS}} & \multicolumn{2}{c|}{\textit{DyARW}}& \multicolumn{3}{c|}{\textit{DyOneSwap}} & \multicolumn{3}{c}{\textit{DyTwoSwap}} \\
		\cline{3-14}
		& &  \textit{gap} & \textit{acc} &  \textit{gap} & \textit{acc} &  \textit{gap} & \textit{acc}  &  \textit{gap} & \textit{acc} & \textit{gap*} &  \textit{gap} & \textit{acc} & \textit{gap*} \\
		\hline
		Epinions	&	26862	&	384	&	98.57\%	&	384	&	98.57\%	&	62	&	99.77\%	&	62	&	99.77\%	 & 24 &	16	&	99.94\% & 3	\cr
		Slashdot	&	30497	&	461	&	98.49\%	&	469	&	98.46\%	&	110	&	99.64\%	&	110	&	99.64\%	& 63 &	34 &	99.89\%	& 18 \cr
		Email	&	199909	&	67	&	99.97\%	&	67	&	99.97\%	&	15	&	99.99\%	&	15 &	99.99\%	& 13 &	2 &	99.99\% & 0	\cr
		com-dblp	&	144175	&	840	&	99.42\%	&	458	&	99.68\%	&	179	&	99.88\%	&	168	&	99.88\% & 126	&	32	&	99.98\% & 18	\cr
		com-amazon	&	160215	&	1130	&	99.29\%	&	860	&	99.46\%	&	623	&	99.61\%	&	630 &	99.61\%	& 465 &	229	&	99.86\%	& 159 \cr
		web-Google	&	506183	&	885	&	99.83\%	&	627	&	99.88\%	&	400	&	99.92\%	&	403 &	99.92\%	& 318 &	152 &	99.97\%	& 128 \cr
		web-BerkStan	&	387092	&	2271	&	99.41\%	&	2071	&	99.46\%	&	2801	&	99.28\%	&	2797 &	99.28\%	& 2183 &	1928 &	99.50\%	& 1488 \cr
		in-2004	&	871575	&	1790	&	99.79\%	&	1593	&	99.82\%	&	2228	&	99.74\%	&	2228 &	99.74\%	& 1841 &	1540 &	99.82\%	& 1310 \cr
		as-skitter	&	1142226	&	317	&	99.97\%	&	245	&	99.98\%	&	711	&	99.94\%	&	711	&	99.94\% & 612	&	255	&	99.98\%	& 228 \cr
		hollywood	&	334268	&	4578	&	98.63\%	&	3699	&	98.89\%	&	29	&	99.99\%	&	32	&	99.99\%	& 28 &	1 &	99.99\%	& 1\cr
		WikiTalk	&	2276357	&	5	&	99.99\%	&	5	&	99.99\%	&	11	&	99.99\%	&	11	&	99.99\%	& 8 &	2	&	99.99\%	& 0 \cr
		com-lj	&	2069002	&	563	&	99.97\%	&	327	&	99.98\%	&	1127	&	99.95\%	&	1127 &	99.95\%	& 889 &	577 &	99.97\%	& 460\cr
		soc-LiveJournal	&	2613955	&	453	&	99.98\%	&	254	&	99.99\%	&	1042	&	99.96\%	&	1041 &	99.96\% & 842	&	523	&	99.98\% & 338	\cr
		\bottomrule
	\end{tabular}\vspace{-4mm}
\end{table*}

\noindent{\bf Evaluate Solution Quality.}
We first evaluate the effectiveness of the proposed algorithms against the existing methods.
We report the gap of the size of the independent set maintained by each algorithm to the independence number computed by {\small \sf VCSolver}~\cite{DBLP:journals/tcs/AkibaI16} and the accuracy achieved by each algorithm after 100,000 updates on thirteen easy real graphs in Table \ref{tab:easy-gap}.
It is clear that not only {\small\it DyTwoSwap} but also {\small\it DyOneSwap} outperforms the competitors {\small\it DGOneDIS} and {\small\it DGTwoDIS} on the first six graphs and achieves a competitive accuracy on the remaining graphs.
As stated previously, in practice, sometimes the amount of updates is quite huge, even equals to the number of vertices in the graph.
Hence, we report the gap and the accuracy of the solution maintained by each algorithm after 1,000,000 updates on the last seven easy graphs in Table \ref{tab:easy-gap2}.
Our methods achieve smaller gaps and higher accuracy on all of them, especially in web-BerkStan and hollywood, with an improvement of 2\% and 4\%, respectively.
The reason is that with the increasing of the number of updates, the competitors fails in more and more rounds to find the set of complementary vertices to avoid the degradation of the solution quality.
And there is no theoretical guarantee on the quality of the maintained solution.

Then, we report the gap of the size of the independent set maintained by each algorithm after 1,000,000 updates to the size of the solution returned by {\small \sf ARW}~\cite{DBLP:journals/heuristics/AndradeRW12} on the hard graphs in Table \ref{tab:hard-gap}.
Notice that {\small\it DGOneDIS} and {\small\it DGTwoDIS} don't finish within five hours on the last five graphs, which is absolutely unacceptable in practice.
The proposed algorithms sometimes even return a solution with more vertices (marked with $\uparrow$).
Although there is no improvement on the approximation ratio, {\small\it DyTwoSwap} is indeed much more effective than {\small\it DyOneSwap} on all graphs.
As for {\small\it DyARW}, since the solution maintained by it is also 1-maximal, its performance is almost the same as {\small\it DyOneSwap} on all graphs.
Therefore, we conclude that our algorithms are more effective especially when the graph is frequently updated, which is quite common in real-life applications.

\begin{table*}[htb]
	\centering
	\caption{The gap to the independence number obtained by {\sf VCSolver}~\cite{DBLP:journals/tcs/AkibaI16} and accuracy on the last seven easy graphs after 1,000,000 updates.}\vspace{-2ex}
	\label{tab:easy-gap2}
	\begin{tabular}{c|c|cc|cc|cc|ccc|ccc}
		\toprule
		\multirow{2}{*}{Graphs} & \multirow{2}{*}{$\alpha(G)$}& \multicolumn{2}{c|}{\textit{DGOneDIS}} & \multicolumn{2}{c|}{\textit{DGTwoDIS}} &  \multicolumn{2}{c|}{\textit{DyARW}} & \multicolumn{3}{c|}{\textit{DyOneSwap}} & \multicolumn{3}{c}{\textit{DyTwoSwap}} \\
		\cline{3-14}
		& &  \textit{gap} & $acc$ &  \textit{gap} & $acc$ &  \textit{gap} & $acc$ &  \textit{gap} & $acc$ & \textit{gap*} &  \textit{gap} & $acc$ & \textit{gap*} \\
		\hline
		web-BerkStan	&	201515	&	6256	&	96.90\%	&	5976	&	97.03\%	&	1302	&	99.35\%	&	1296 &	99.36\%	& 827 &	498	&	99.75\%	& 379 \cr
		in-2004	&	656141	&	11151	&	98.30\%	&	10024	&	98.47\%	&	3511	&	99.46\%	&	3519 &	99.46\%	& 2035 &	1093 &	99.83\%	& 421 \cr
		as-skitter	&	903919	&	6348	&	99.30\%	&	5912	&	99.35\%	&	3200	&	99.65\%	&	3203 &	99.65\% & 2568	&	1076	&	99.88\% &	198 \cr
		hollywood	&	351317	&	17845	&	94.92\%	&	13419	&	96.18\%	&	1020	&	99.71\%	&	1030 &	99.71\%	& 748 &	141	&	99.96\% &	69 \cr
		WikiTalk	&	1802293	&	521	&	99.97\%	&	518	&	99.97\%	&	104	&	99.99\%	&	104	&	99.99\%	& 83 &	9	&	99.99\%	& 6 \cr
		com-lj	&	1918084	&	11207	&	99.42\%	&	9911	&	99.48\%	&	7496	&	99.61\%	&	7500	&	99.61\% & 5351	&	3436	&	99.82\%	& 2481 \cr
		soc-LiveJournal	&	2452504	&	9690	&	99.60\%	&	8205	&	99.67\%	&	7066	&	99.71\%	&	7052	&	99.71\%	& 5971 &	3096	&	99.87\%	& 2206 \cr
		\bottomrule
	\end{tabular}
	\vspace{-4mm}
\end{table*}

\begin{table*}[htb]
	\centering
	%\fontsize{7}{8.5}\selectfont
	\caption{The gap to the best result size obtained by the local search algorithm {\sf ARW}~\cite{DBLP:journals/heuristics/AndradeRW12} on hard graphs after 1,000,000 updates}
	\vspace{-2ex}
	\label{tab:hard-gap}
	\begin{tabular}{c|c|ccccc}
		\toprule
		\multirow{2}{*}{Graphs} & \multirow{2}{*}{Best Result}& \multicolumn{4}{c}{\textit{Gap to the Best Result Size}}\\
		\cline{3-7}
		&  &\multicolumn{1}{c|}{\it DGOneDIS} & \multicolumn{1}{c|}{\it DGTwoDIS} & \multicolumn{1}{c|}{\it DyARW} &  \multicolumn{1}{c|}{{\it DyOneSwap} (gap*)} & \multicolumn{1}{c}{{\it DyTwoSwap} (gap*)}\\
		\hline
		soc-pokec	& 612,901 & 4,006 & 3,939 & 1,157$\uparrow$ & 1,143$\uparrow$ (1,272$\uparrow$) & 3,595$\uparrow$ (3,535$\uparrow$) \cr
		wiki-topcats & 792,023 & 10,885 & 10,100 & 4,024 & 4,013 (2,390) & 882 (212)\cr
		com-orkut	& 747,459 & 4,669 & 3,037 & 3,348 & 3,347 (2,557$\uparrow$) & 5,062$\uparrow$ (9,338$\uparrow$)\cr
		cit-Patents	& 1,865,112 & 6,795 & 6,261 & 5,521 & 5,509 (1,825) & 276$\uparrow$ (2,480$\uparrow$)\cr
		uk-2005 & 23,363,561 & - & - & 7,442 & 7,443 (5,164) & 227$\uparrow$ (2,893$\uparrow$)\cr
		it-2004 & 25,238,765 & - & - & 13,083 & 13,078 (8,427) & 276$\uparrow$ (4,517$\uparrow$)\cr
		twitter-2010 & 28,423,449 & - & - & 6,870 & 6,871 (3,742) & 3,515 (142$\uparrow$)\cr
		Friendster & 36,012,590 & - & - & 5,241 & 5,248 (2,929) & 703$\uparrow$ (3,149$\uparrow$)\cr
		uk-2007 & 68,976,197 & - & - & 12,741 & 12,746 (8,354) & 15,974$\uparrow$ (18,291$\uparrow$)\cr
		\bottomrule
	\end{tabular}\vspace{-4mm}
\end{table*}

\begin{figure}[htb]
	\centering
	\subfigure[Response time caused by 100,000 updates]{
		\includegraphics[width=.9\linewidth]{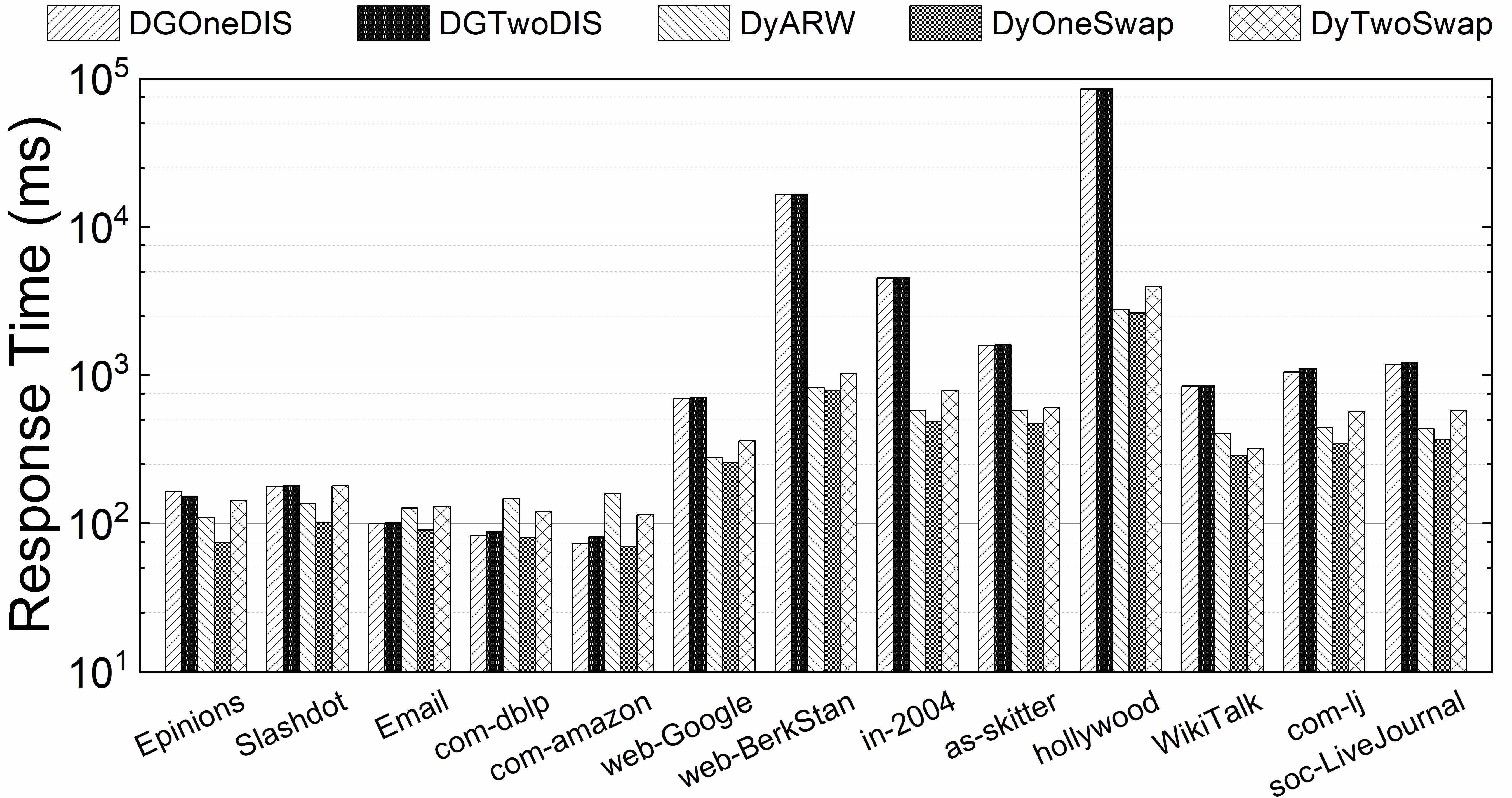}
		\label{fig:easy-eff1}
	}\vspace{-1ex}
	\subfigure[Memory usage caused by 100,000 updates]{
		\includegraphics[width=.9\linewidth]{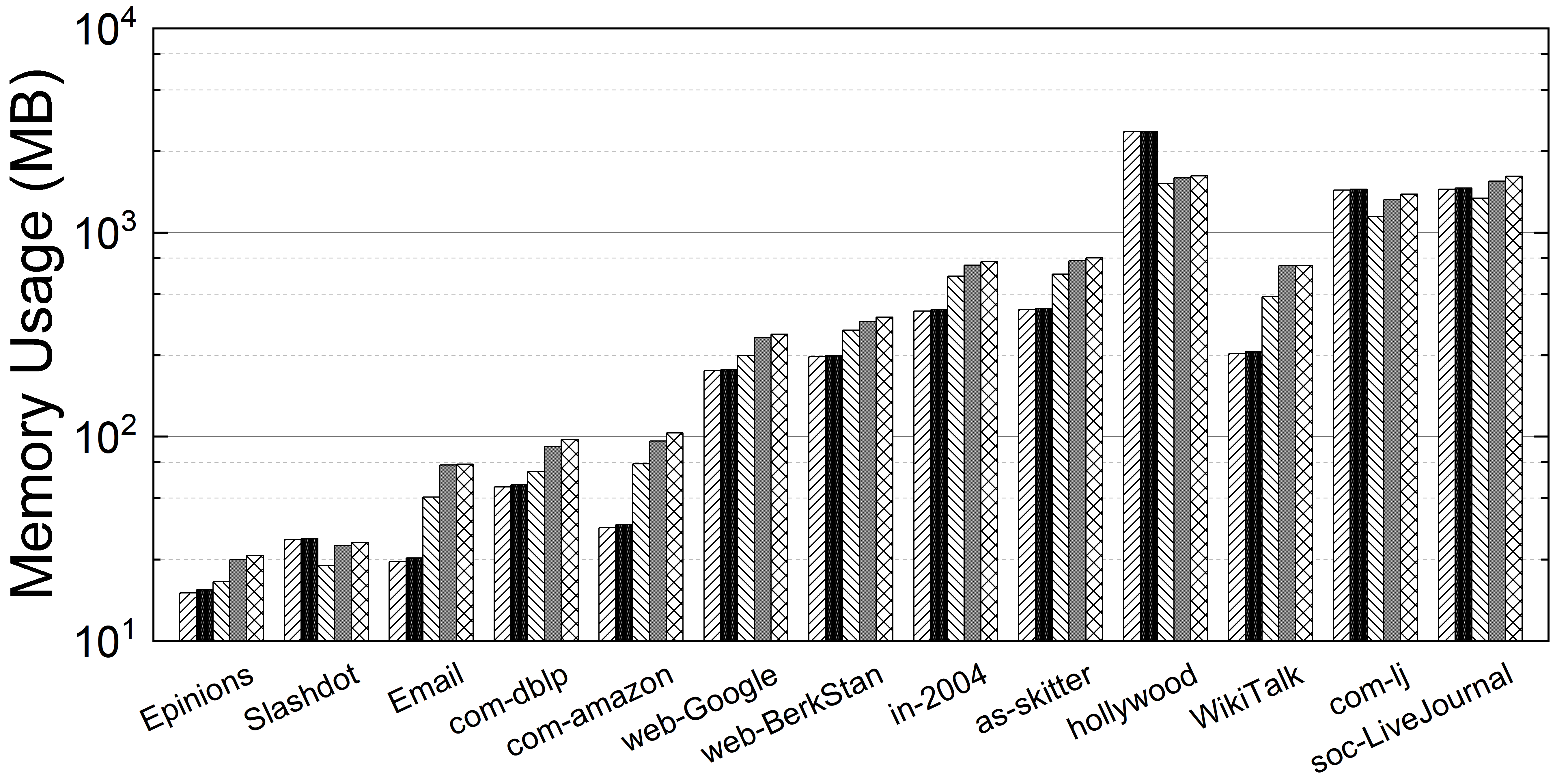}
		\label{fig:easy-mem}
	}\vspace{-1ex}
	\subfigure[Response time caused by 1,000,000 updates]{
		\includegraphics[width=.9\linewidth]{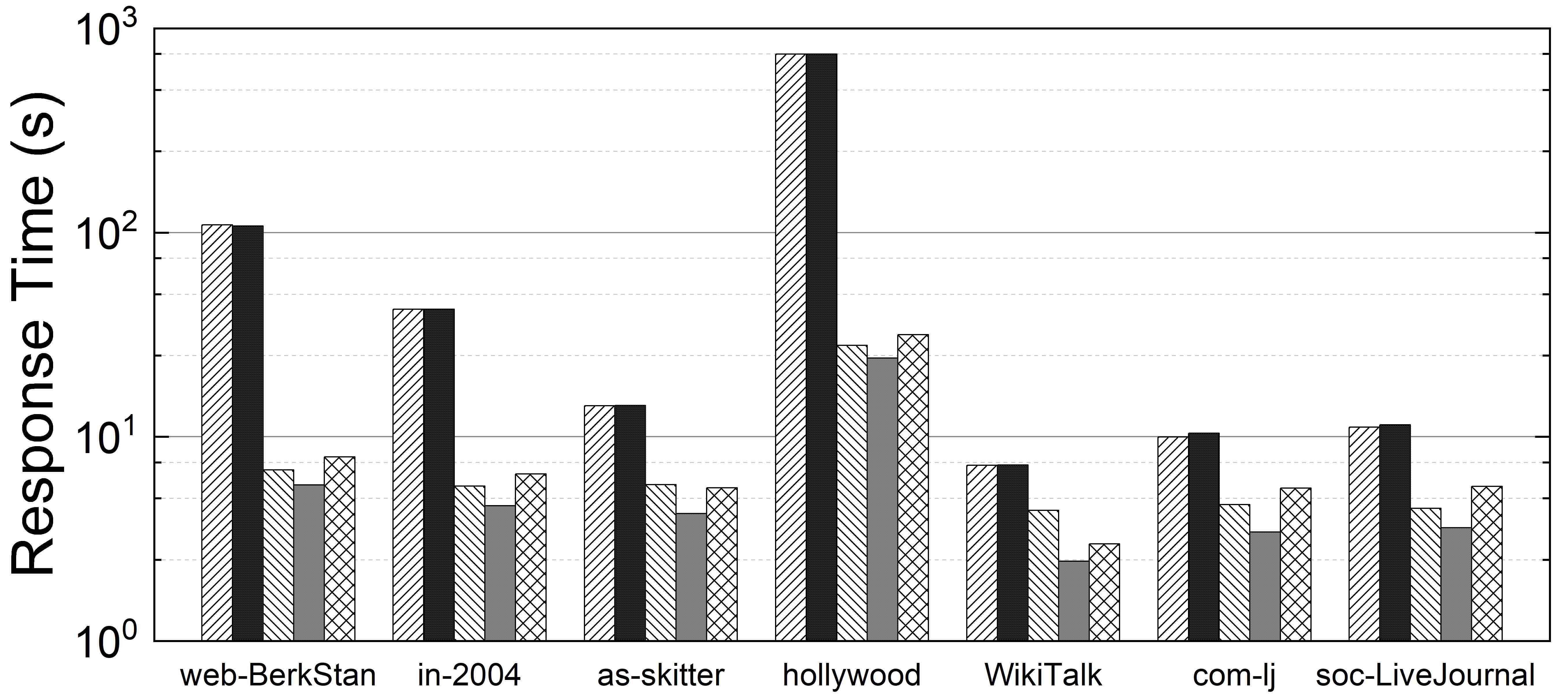}
		\label{fig:easy-eff2}
	}\vspace{-1ex}
	\label{fig:exp-easy}
	\caption{Response time and memory usage on easy graphs}
	\vspace{-2mm}
\end{figure}

\begin{figure}[htb]
	\centering
	\subfigure[Response time caused by 1,000,000 updates]{
		\includegraphics[width=.95\linewidth]{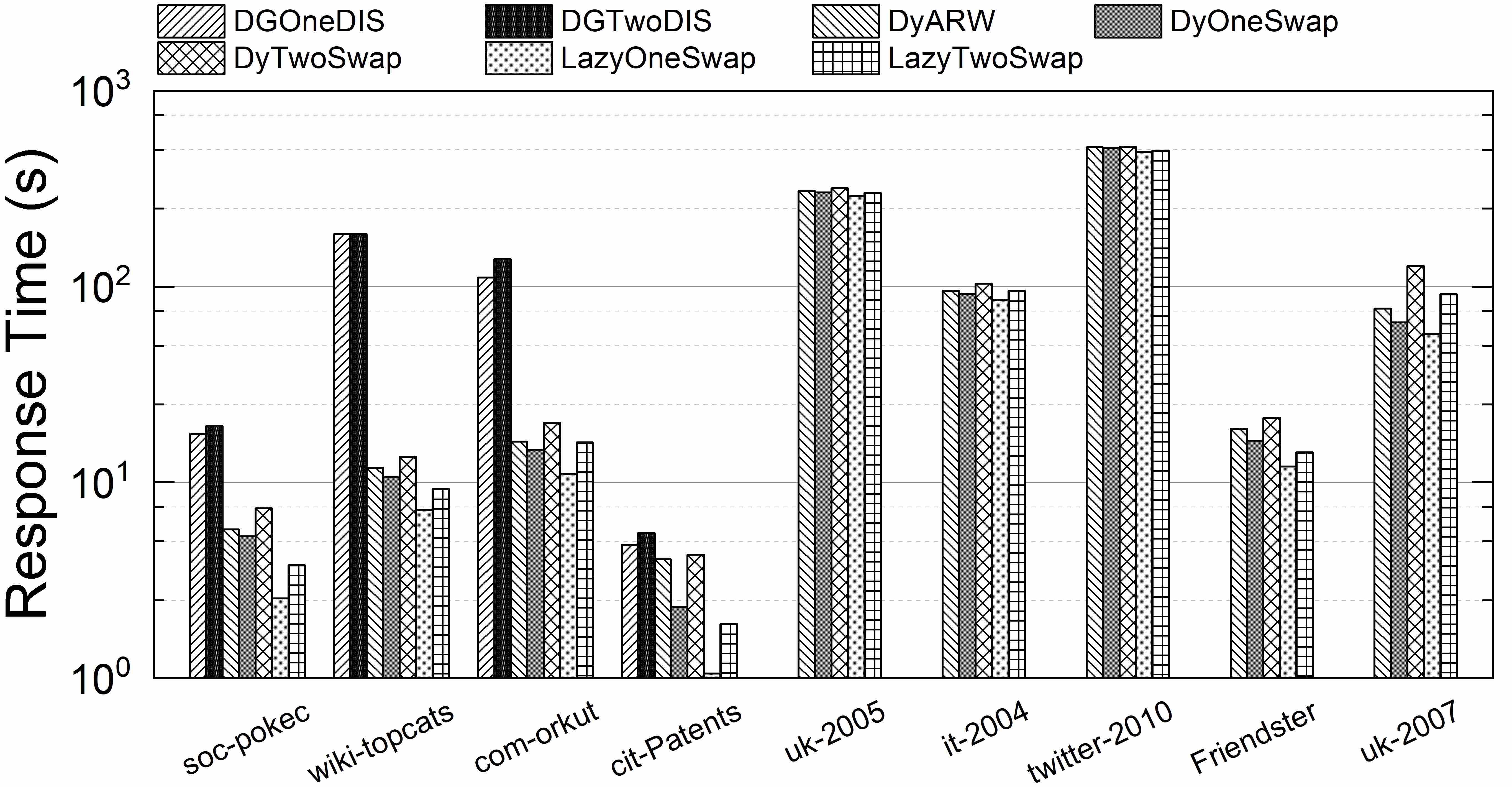}
		\label{fig:hard-eff}
	}\vspace{-1ex}
	\subfigure[Memory usage caused by 1,000,000 updates]{
		\includegraphics[width=.95\linewidth]{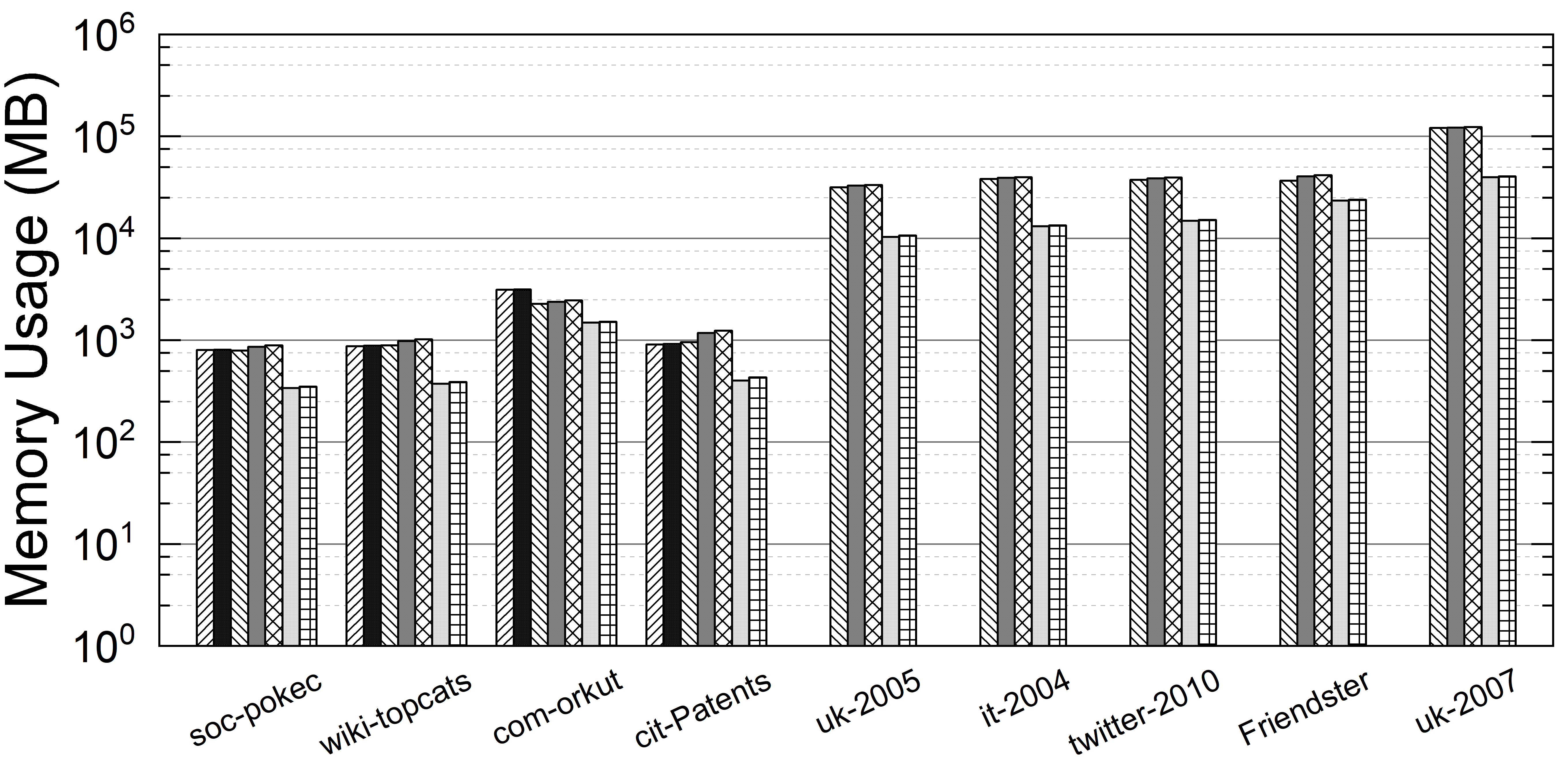}
		\label{fig:hard-mem}
	}
	\label{fig:exp-hard}\vspace{-1ex}
	\caption{Response time and memory usage on hard graphs}
	\vspace{-2mm}
\end{figure}

\noindent{\bf Evaluate Time Efficiency.}
To compare the time efficiency of these algorithms, the time consumed by each of them to process 100,000 updates on the thirteen easy real graphs are shown in Fig. \ref{fig:easy-eff1}.
Generally, the response time of all five algorithms increase along with the increasing of the graph size.
Due to its simplicity, {\small\it DyOneSwap} runs the fastest across all graphs.
Although with the same strategy as {\small\it DyOneSwap}, {\small\it DyARW} suffer from a little higher maintenance time for the ordered structure required by the double pointer scan implementation.
Both {\it\small DyOneSwap} and {\small\it DyTwoSwap} runs faster than {\it\small DGOneDIS} and {\small\it DGTwoDIS} on most of the graphs, especially when the graph is dense \eg, hollywood.
Since 2-swap is additionally considered, {\small\it DyTwoSwap} takes a little more time than {\small\it DyOneSwap}.
We also show the response time taken by each algorithm to handle 1,000,000 updates on the last seven easy graphs and hard graphs in Fig. \ref{fig:easy-eff2} and Fig. \ref{fig:hard-eff}, respectively.
It is surprising that {\it\small DGOneDIS} and {\small\it DGTwoDIS} suffer from a very high time consumption due to the huge search space, especially in web-Berkstan and hollywood.
Moreover, they even didn't finish in five hours on the last five hard graphs.
Considering the performance of {\it\small DGOneDIS} and {\small\it DGTwoDIS} when the number of updates is small, it is noticed that the initial dependency represented by the index is quite simple as it is constructed based on degree-one reduction and degree-two reduction.
However, as the graph evolves, the index becomes more and more complex which leads to a huge search space.
%This explains their high time consumption in frequently updated graphs.

\noindent{\bf Evaluate Memory Usage.}
The memory usage of each algorithm on easy graphs and hard graphs is shown in Fig.~\ref{fig:easy-mem} and Fig.~\ref{fig:hard-mem}, respectively.
The memory usage of all algorithms increase with the increasing of the graph size.
Since {\it \small DyOneSwap} and {\small\it DyTwoSwap} maintain more information to speed up the swap operations and store additional position indices to enable constant-time update of the information, they consume more space than {\small\it DGOneDIS} and {\small\it DGTwoDIS}.
And {\small\it DyTwoSwap} consumes more space than {\small\it DyOneSwap} because it additionally maintains vertices in $\bar{I}_2$ for efficiently identifying 2-swaps.
Since the memory usage of the proposed methods is less than 10GB in most graphs, and does not exceed the maximum available memory of the machine even on large graphs like Friendster and uk-2007, we conclude that the memory consumption is acceptable.
Moreover, we come up an optimization strategy that significantly reduce the memory consumption which is evaluated in the following.

\begin{figure*}
	\centering\hspace{-4ex}
	\subfigure[Response Time]{
		\includegraphics[width=0.23\linewidth]{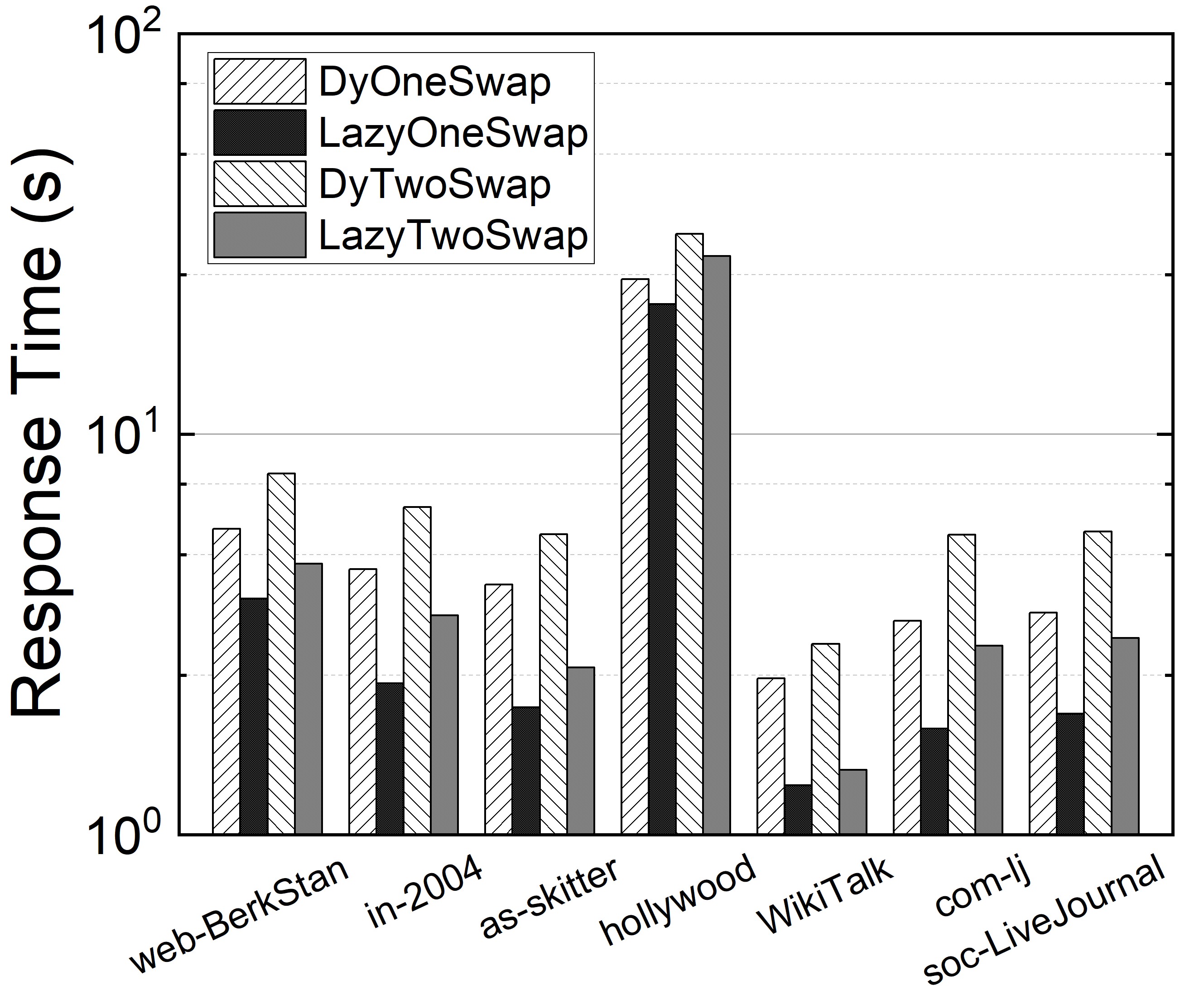}
		\label{fig:lazy-eff}
	}\hspace{-1ex}
	\subfigure[Memory Usage]{
		\includegraphics[width=0.23\linewidth]{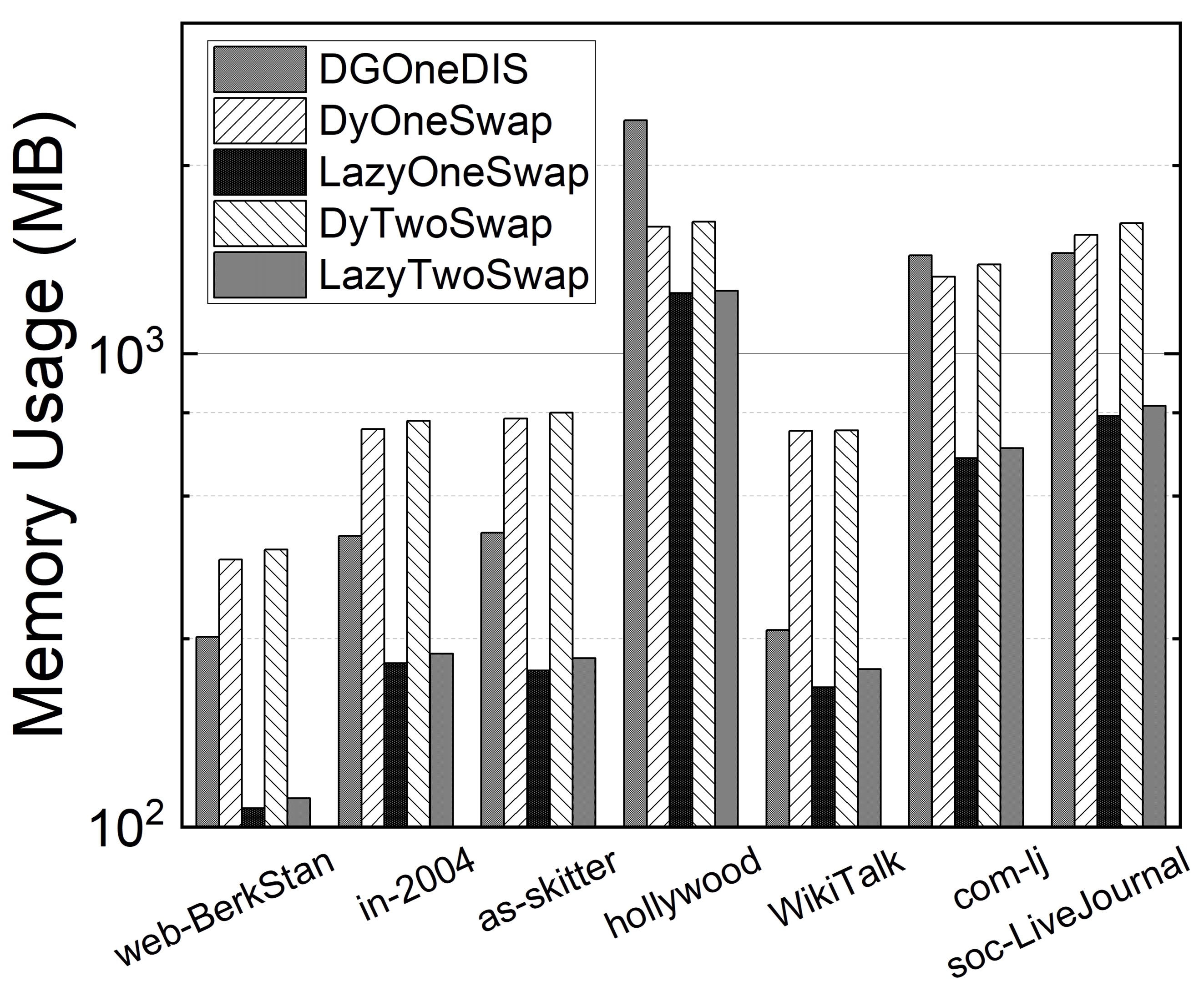}
		\label{fig:lazy-mem}
	}\hspace{-1ex}
	\subfigure[Response Time]{
		\includegraphics[width=0.23\linewidth]{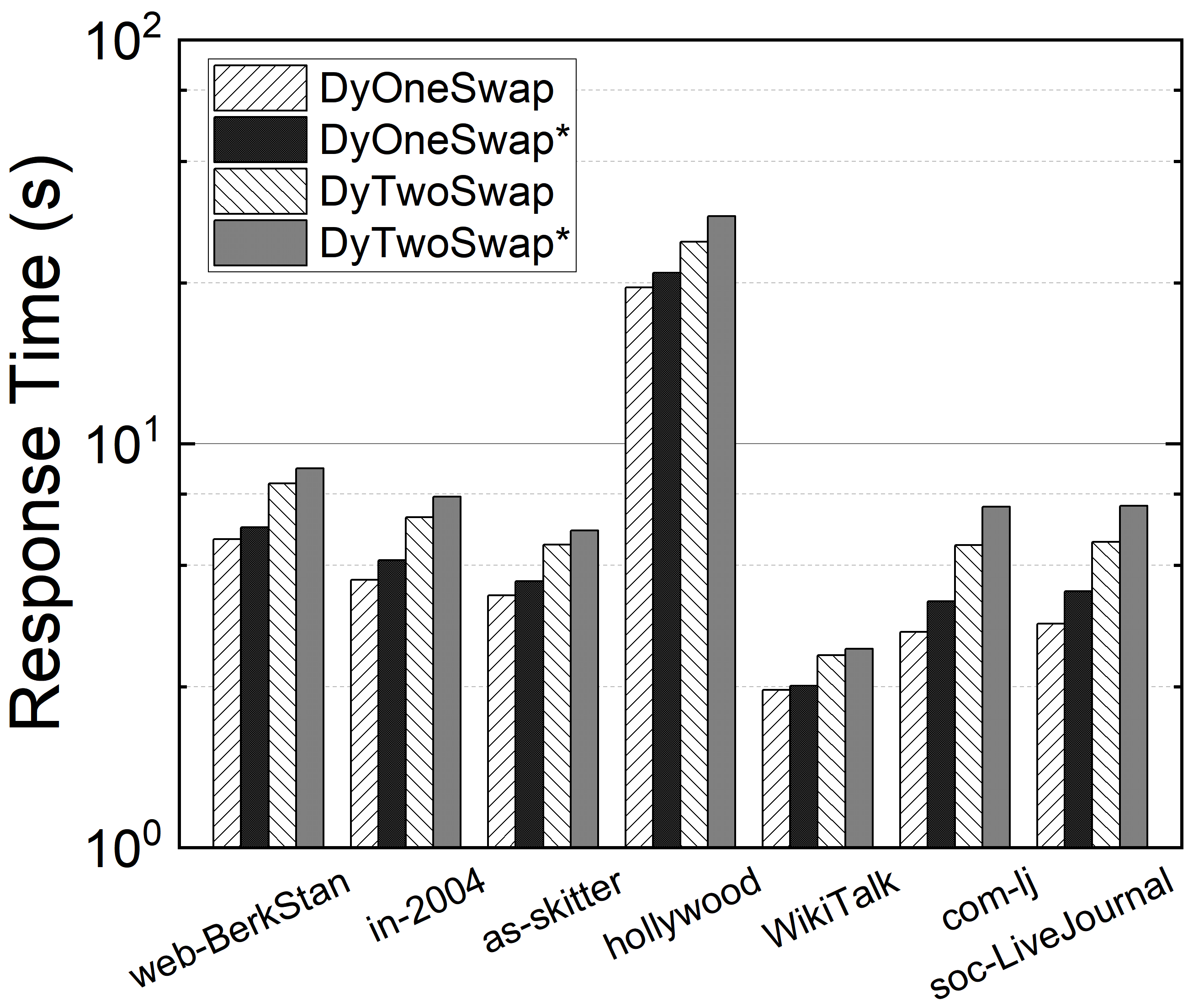}
		\label{fig:op-eff}
	}\hspace{-1ex}
	\subfigure[Time Improvement]{
		\includegraphics[width=0.235\linewidth]{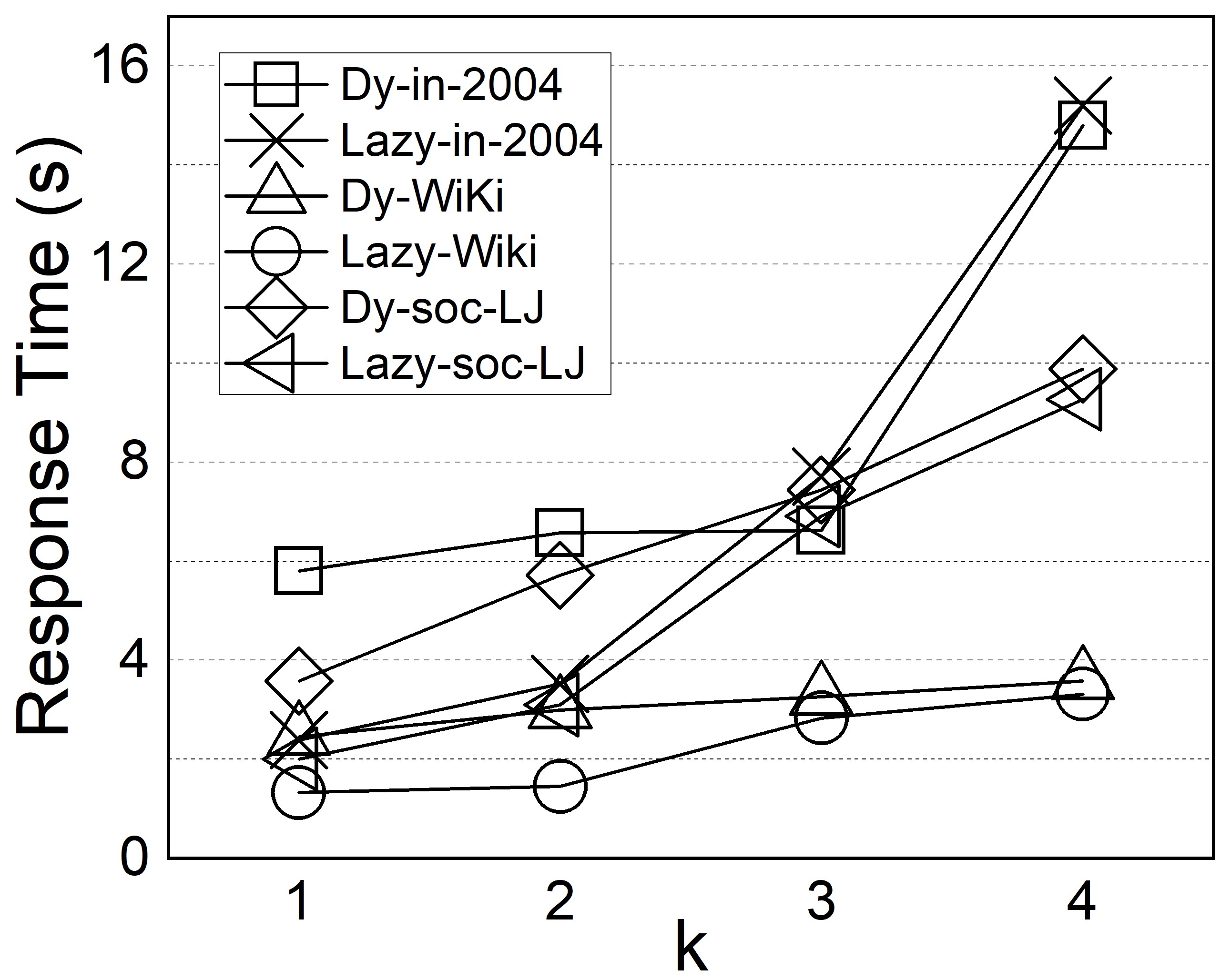}
		\label{fig:scal-k-improve}
	}\vspace{-1ex}
	\caption{Evaluation of Optimizations.}\vspace{-2mm}
	\label{fig:lazy}\vspace{-2mm}
\end{figure*}

\noindent{\bf Evaluate Optimizations.}
We first evaluate the effect of {\it lazy collection strategy} on the response time and memory usage of the proposed algorithms.
As shown in Fig.~\ref{fig:lazy-mem} and Fig.~\ref{fig:hard-mem}, the memory consumption is significantly reduced due to the fact that only $count$ for each vertex is maintained in the algorithm.
Moreover, this strategy also helps to improve time efficiency when $k$ is small.
But, as indicated in Fig.~\ref{fig:scal-k-improve}, the time consumption goes higher as $k$ increases, which indicates an interesting trade-off between the maintenance time and the calculation time under the dynamic setting.
Then, we evaluate the effect of {\it perturbation} on the quality of the solution maintained by the proposed algorithms.
We report the gap achieved by each algorithm equipped with {\it perturbation} in the {\it gap*} column of Table~\ref{tab:easy-gap}, Table~\ref{tab:easy-gap2}, and Table~\ref{tab:hard-gap}.
Even though the original gap achieved by each algorithm is already small, there is still improvement by using {\it perturbation} with a little higher time consumption as shown in Fig.~\ref{fig:op-eff}.

\begin{figure*}[htb]
	\centering
	\subfigure[Response Time (hollywood)]{
		\includegraphics[width=0.225\linewidth]{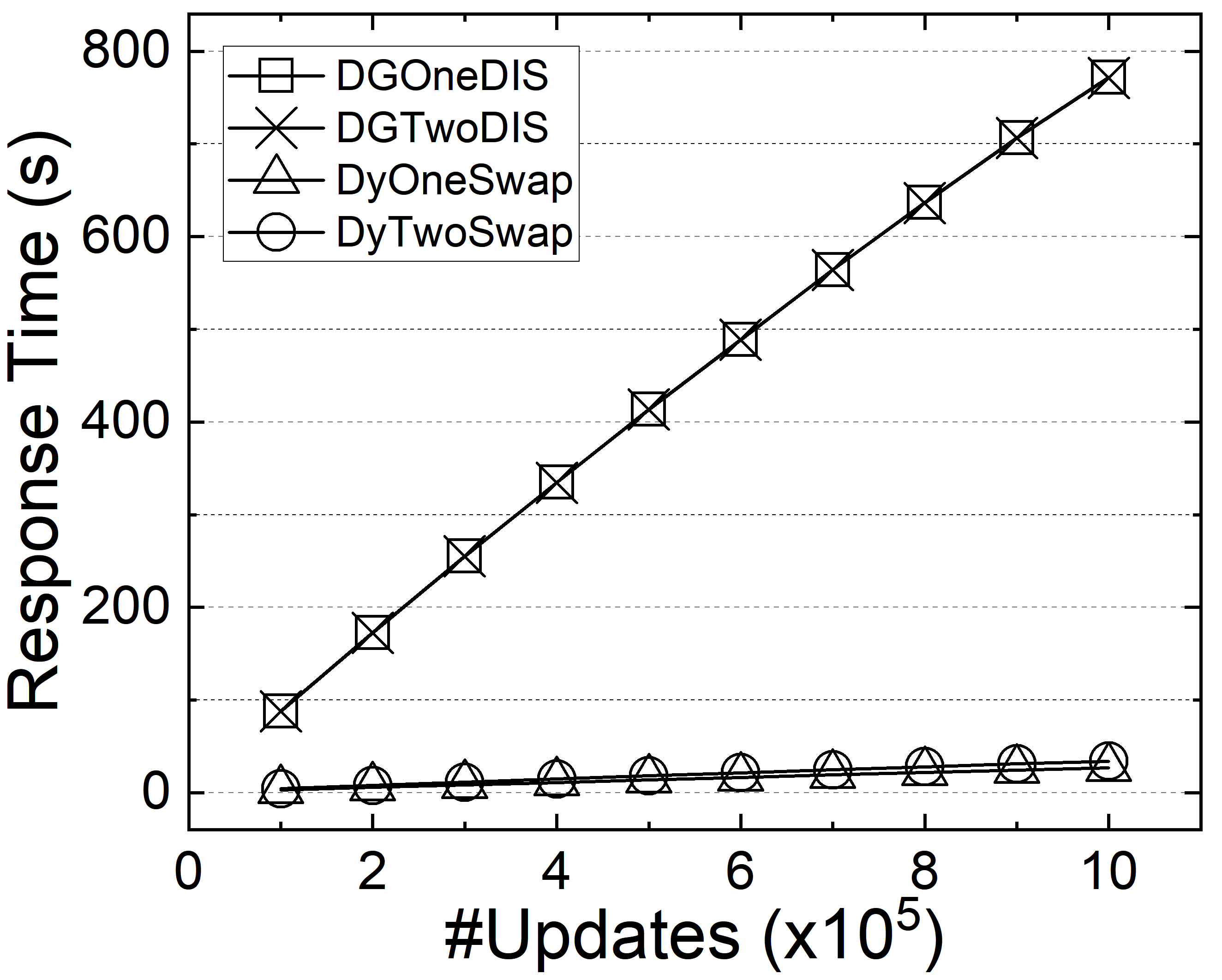}
		\label{fig:scal-eff-l}
	}\hspace{-1ex}
	\subfigure[Gap\&Accuracy (hollywood)]{
		\includegraphics[width=0.25\linewidth]{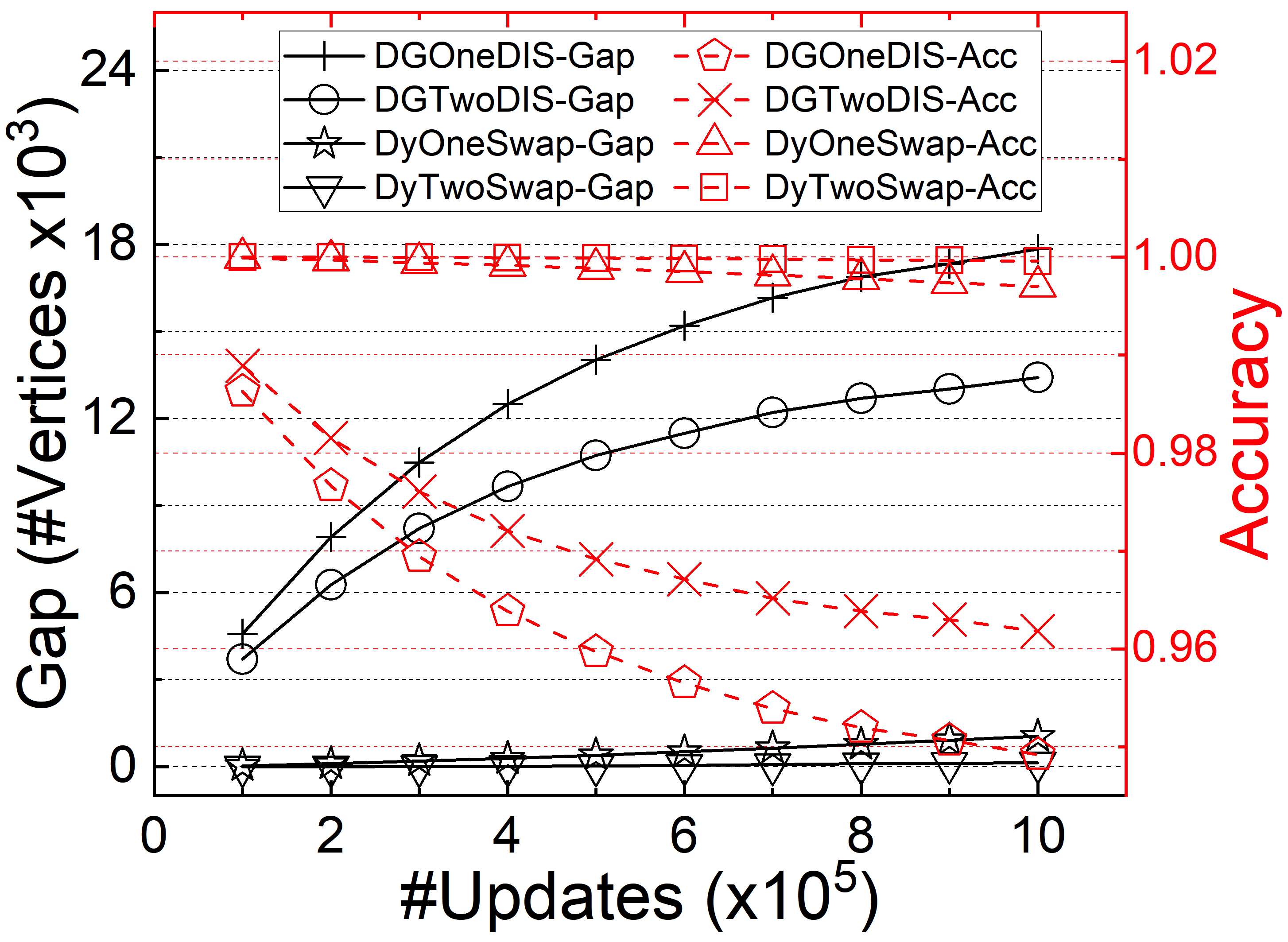}
		\label{fig:scal-gap-l}
	}\hspace{-1ex}
	\subfigure[Response Time (soc-LiveJournal)]{
		\includegraphics[width=0.22\linewidth]{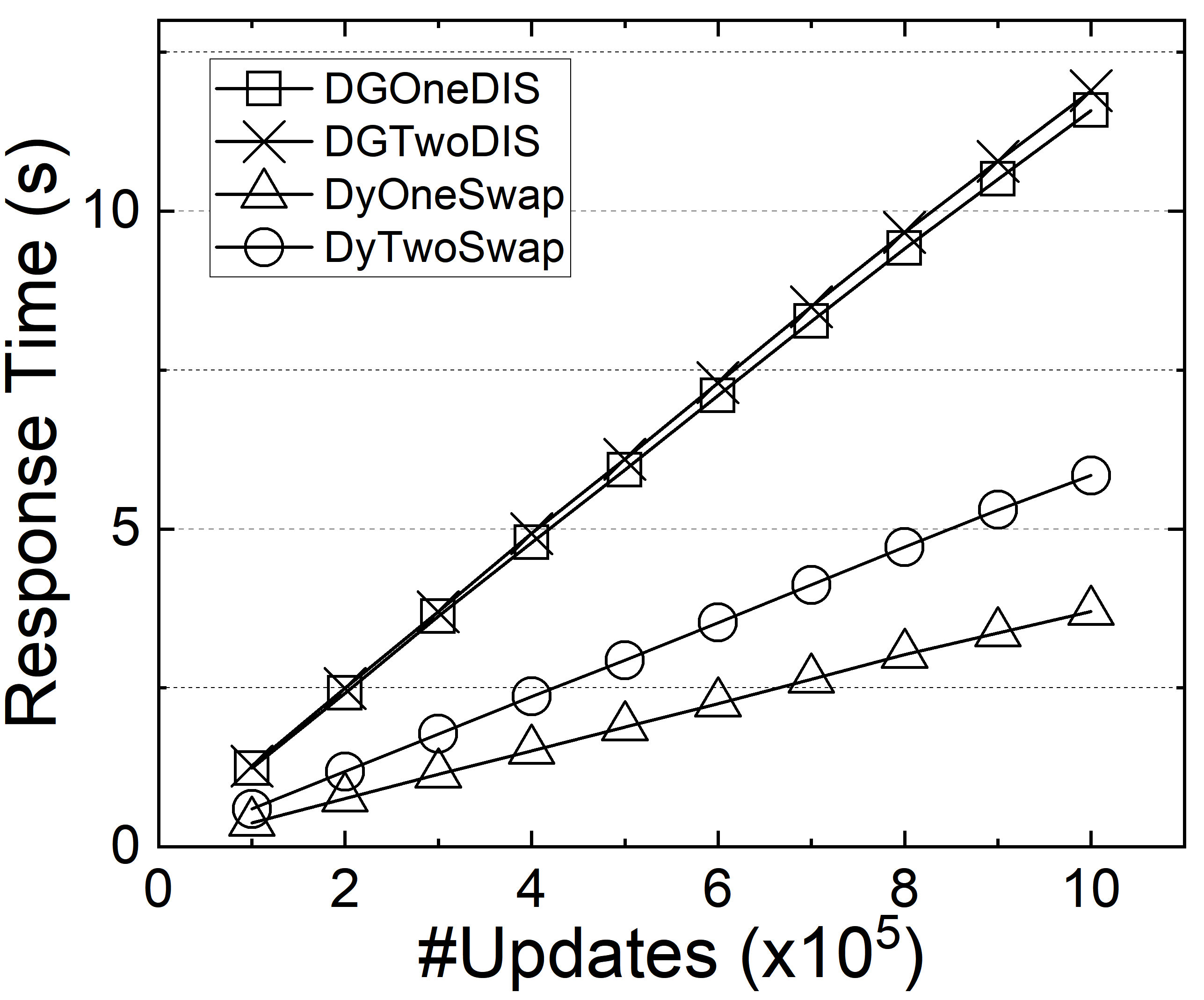}
		\label{fig:scal-eff-b}
	}\hspace{-1ex}
	\subfigure[Gap\&Accuracy (soc-LiveJournal)]{
		\includegraphics[width=0.25\linewidth]{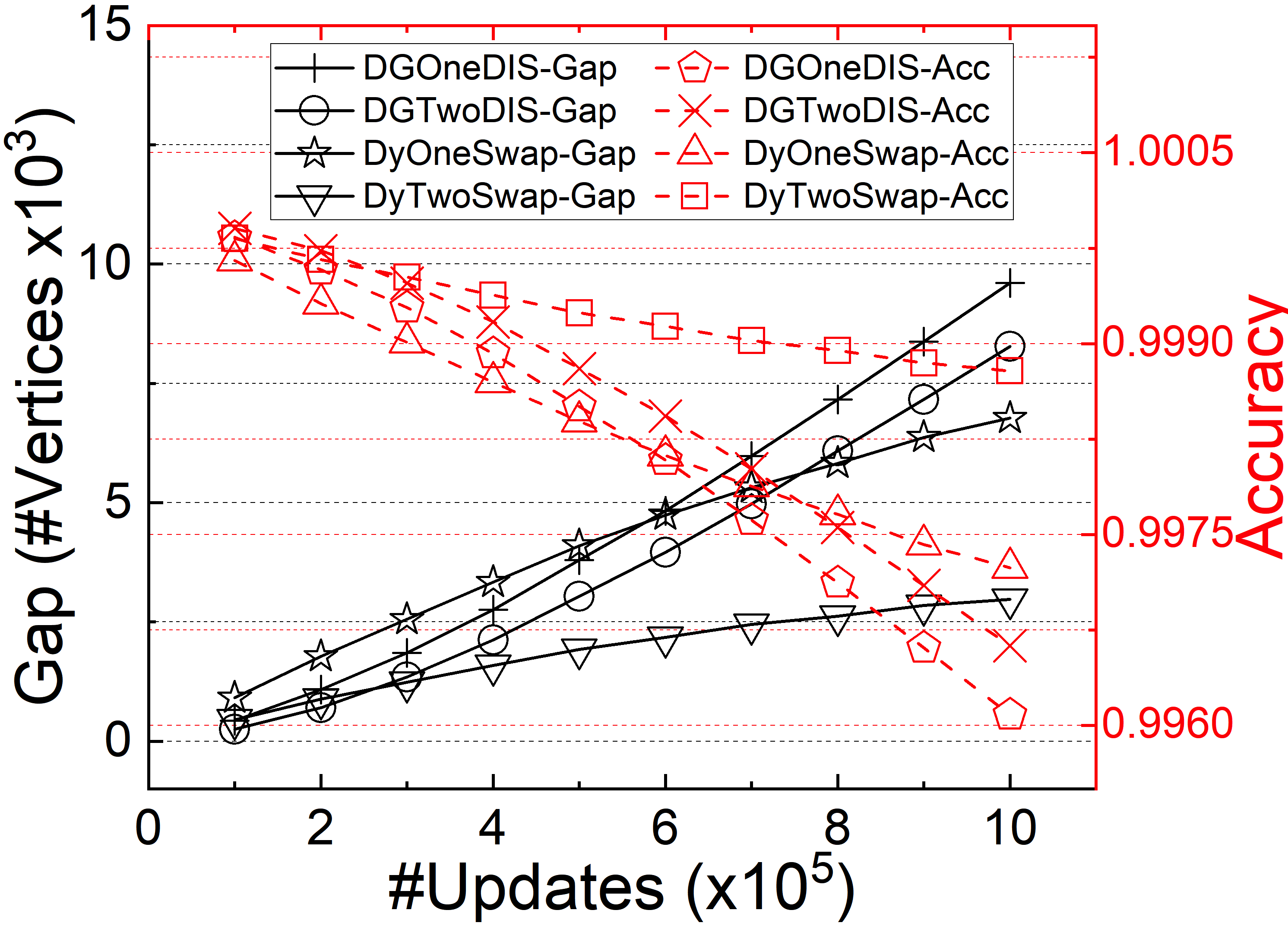}
		\label{fig:scal-gap-b}
	}\vspace{-1ex}
	\caption{Scalability evaluation on hollywood and soc-LiveJournal}\vspace{-2mm}
	\label{fig:scal}\vspace{-2mm}
\end{figure*}

\begin{figure}
	\centering
	\subfigure[Response Time]{
		\includegraphics[width=0.45\linewidth]{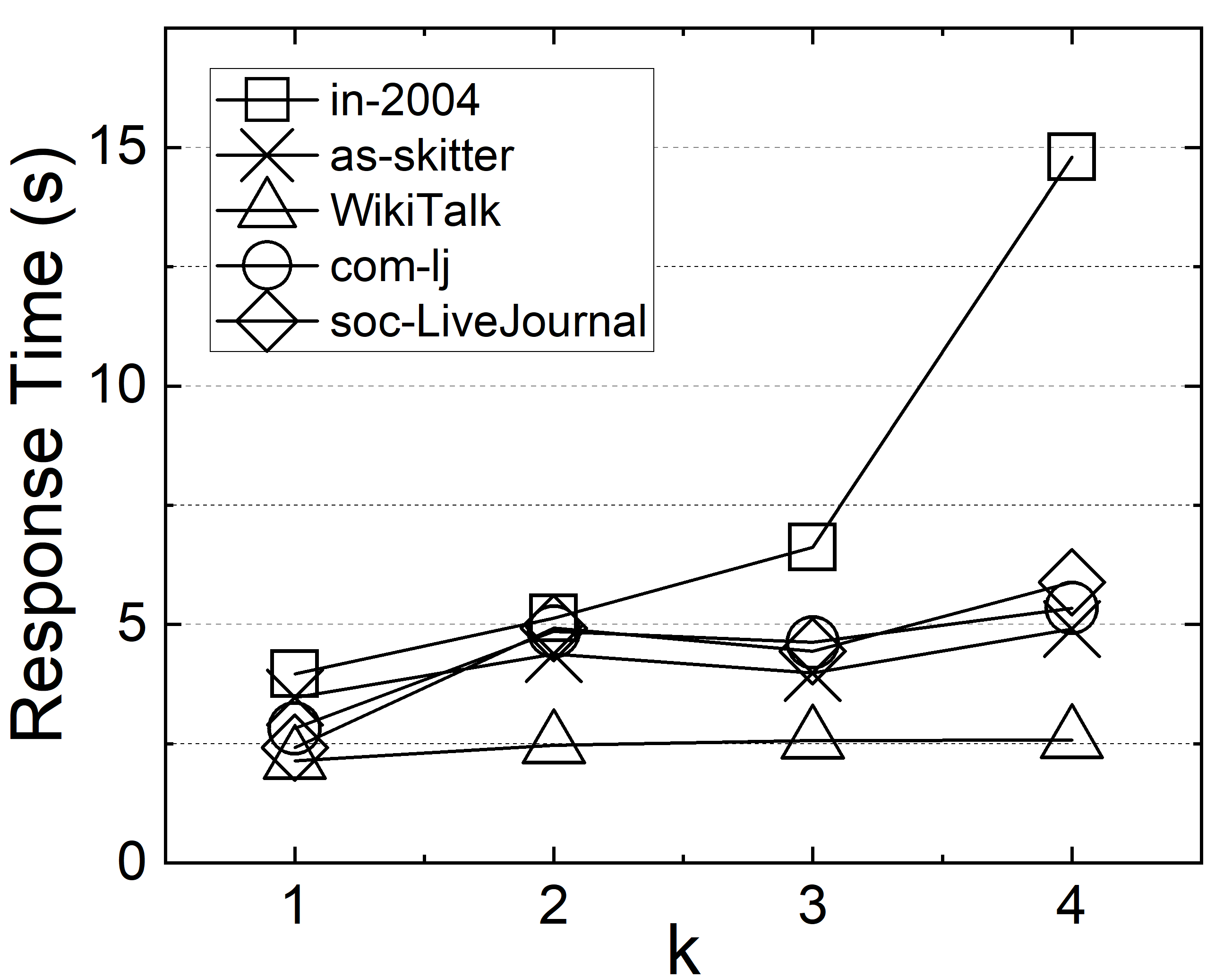}
		\label{fig:scal-k-eff}
	}\hspace{-2ex}
	\subfigure[Gap\&Accuracy]{
		\includegraphics[width=0.515\linewidth]{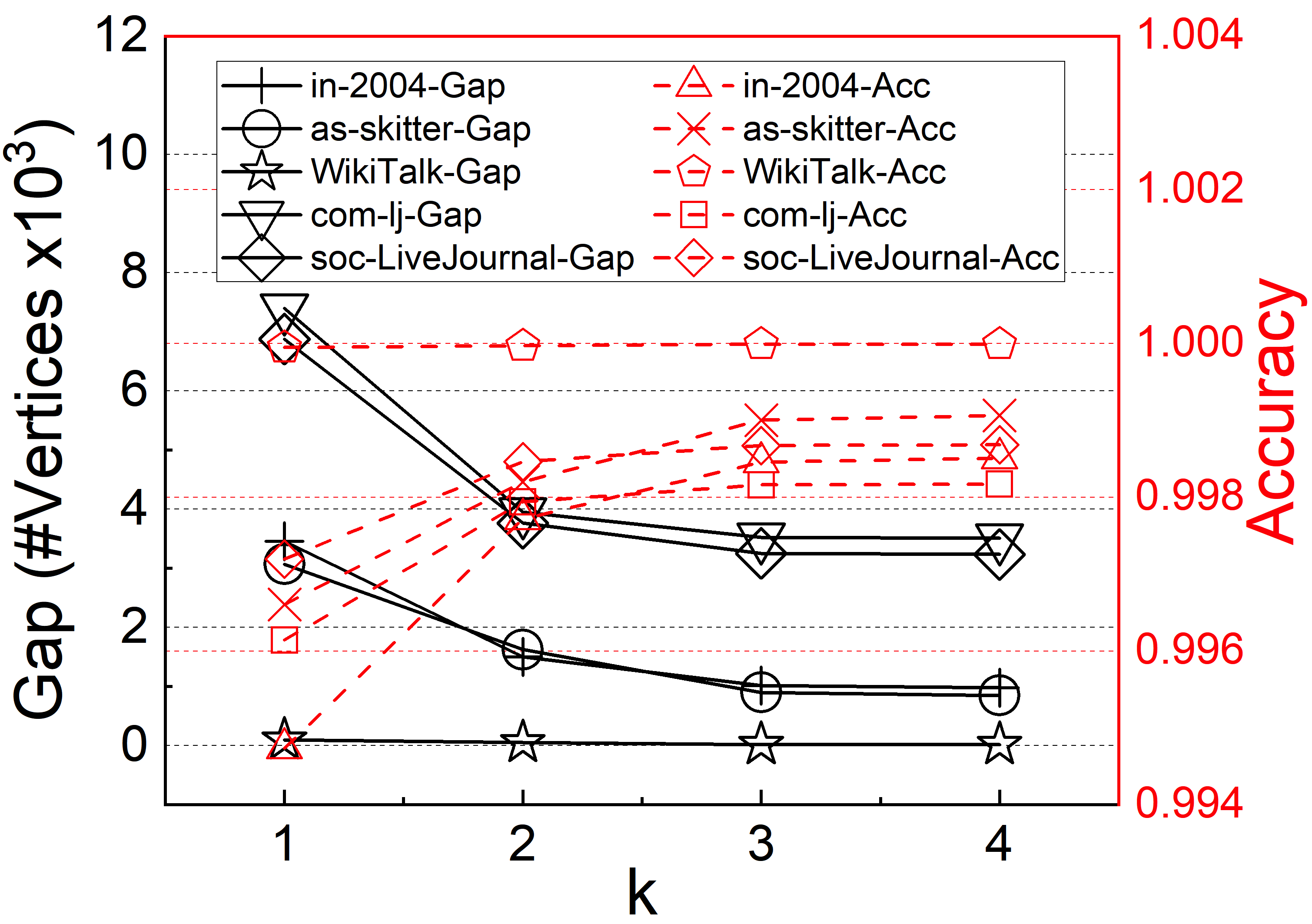}
		\label{fig:scal-k-acc}
	}\vspace{-1ex}
	\caption{Scalability evaluation of $k$}\vspace{-2mm}
	\label{fig:scal-k}\vspace{-2mm}
\end{figure}

\noindent{\bf Evaluate Scalability.}
To study the scalability of the proposed algorithms, we first vary the number of update operations (denoted by \#Updates) from 100,000 to 1,000,000, and plot the performance of each algorithm in hollywood and soc-LiveJournal.
Fig.~\ref{fig:scal-eff-l} and Fig.~\ref{fig:scal-eff-b} show the effect of \#Updates on the time efficiency.
It is clear that the increasing rate of the response time is near linear to the amount of update operations.
And, the improvement of {\it\small DyTwoSwap} and {\it\small DyOneSwap} in time efficiency is stable and significant, especially in hollywood.
Fig.~\ref{fig:scal-gap-l} and Fig.~\ref{fig:scal-gap-b} show the effect of \#Updates on the gap and accuracy.
As we can see, the performance of all algorithms degrades with the number of updates increases.
However, the proposed methods have a lower decreasing rate than the competitors.
Then, we evaluate the effect of $k$ on the time efficiency and the accuracy.
As shown in Fig.~\ref{fig:scal-k-eff} and Fig.~\ref{fig:scal-k-acc}, a larger $k$ means higher solution quality but also higher time consumption.
Therefore, when setting $k$ for a real-world application, it mainly depends on the update frequency of the underlying graph.
The higher the frequency, the smaller the recommended $k$.
The accuracy is also well guaranteed even when $k = 1$.

\begin{figure}[htb]
\centering
\subfigure[Response Time]{
	\includegraphics[width=0.45\linewidth]{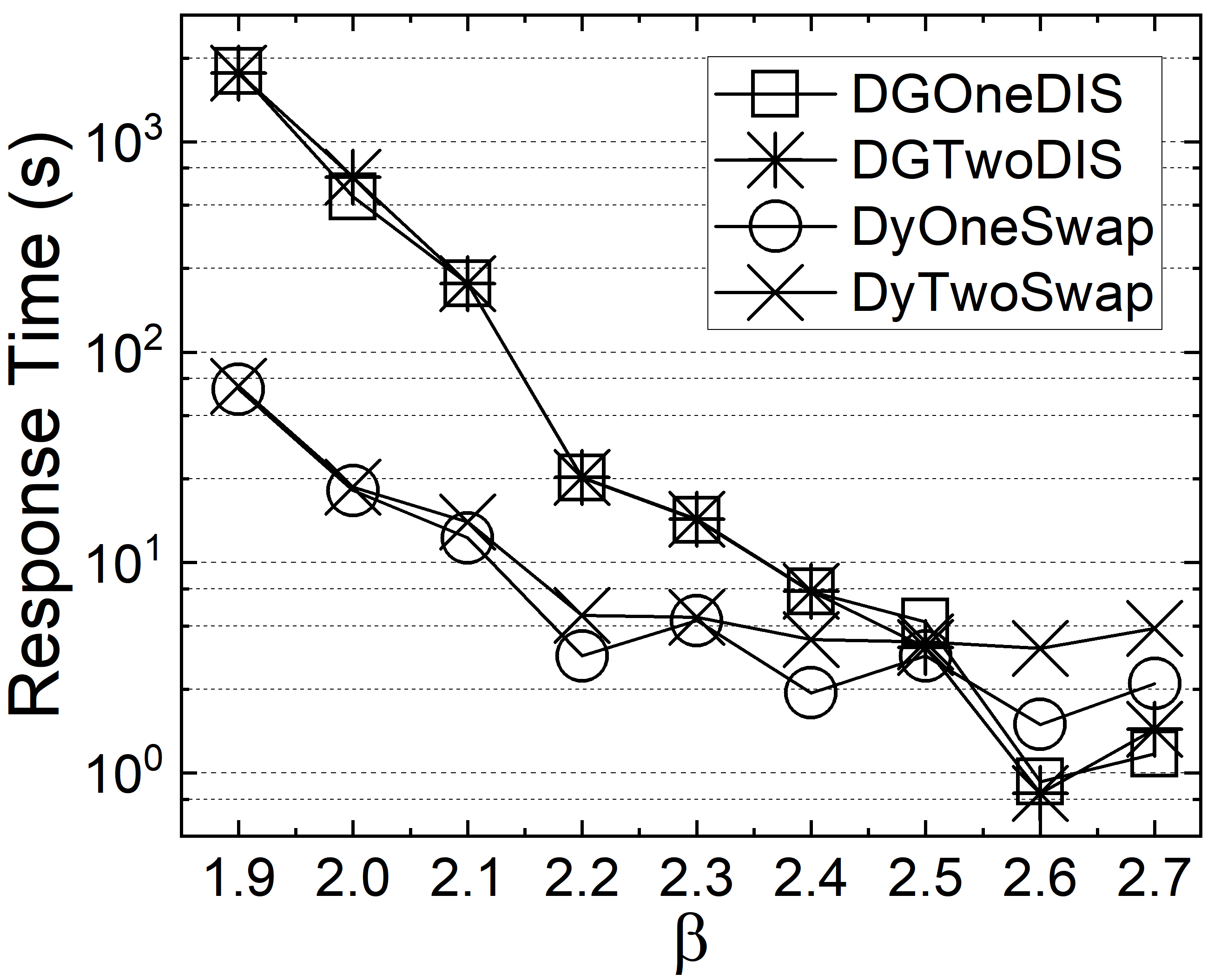}
	\label{fig:plr-eff}
}\hspace{-2ex}
\subfigure[Gap\&Accuracy]{
	\includegraphics[width=0.51\linewidth]{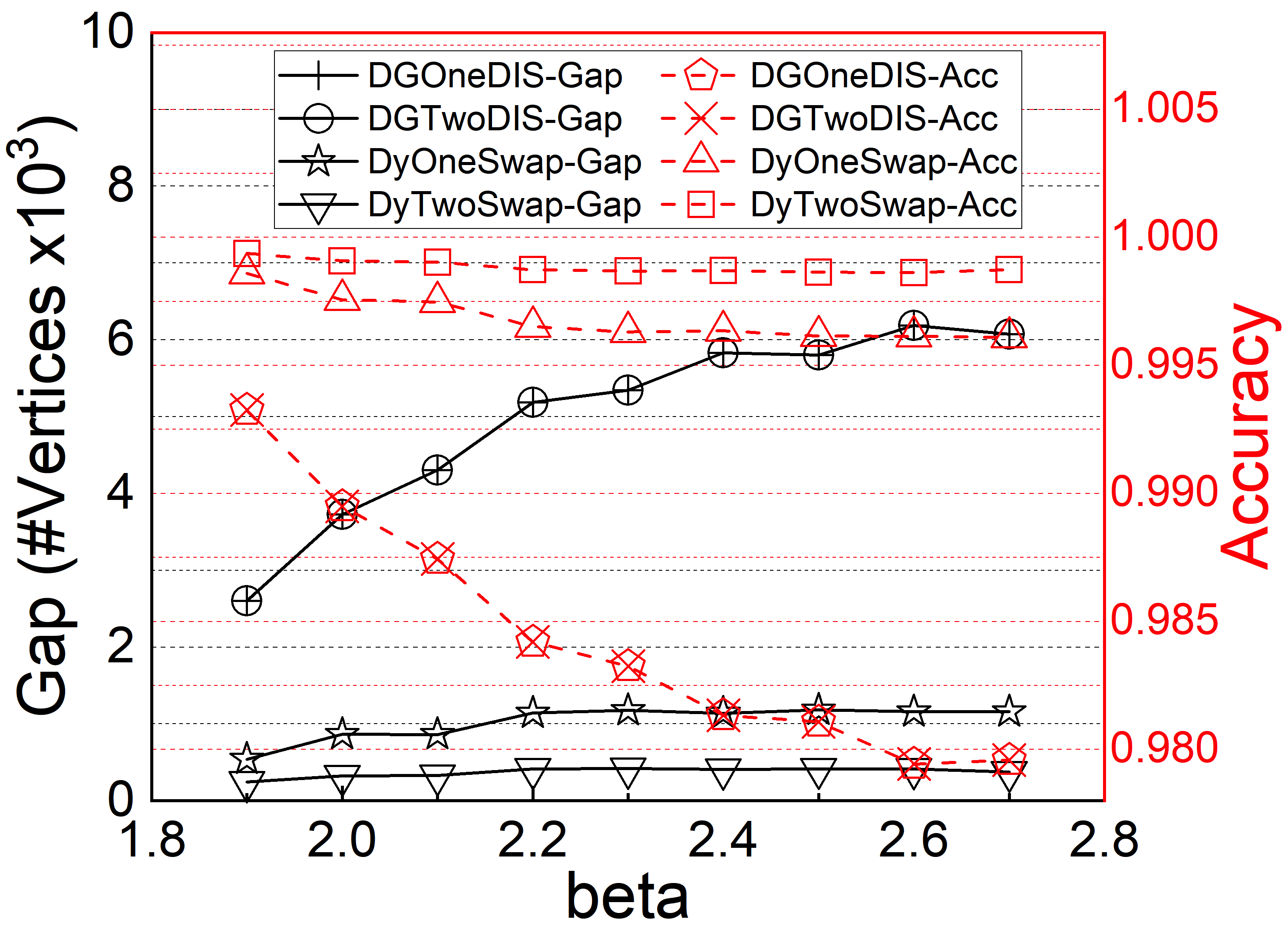}
	\label{fig:plr-acc}
}\vspace{-1ex}
\caption{Performance on Power-Law Random Graphs}\vspace{-2mm}
\label{fig:plr}\vspace{-2mm}
\end{figure}

\noindent\underline{\it Power-Law Graphs.}
We generate nine Power-Law Random (PLR) graphs using NetworkX\footnote{http://networkx.github.io/} with $10^6$ vertices by varying the growth exponent $\beta$ from $1.9$ to $2.7$.
The results on these PLR graphs are shown in Fig. \ref{fig:plr}.
It is easy to see that the proposed methods {\it\small DyOneSwap} and {\small\it DyTwoSwap} outperform the competitors {\small\it DGOneDIS} and {\it\small DGTwoDIS} significantly in terms of both gap (accuracy) and response time.
The proposed methods are better than the competitors by a margin around 1.5\% when $\beta$ is small, which is a noticeable improvement.
Moreover, both {\it\small DGOneDIS} and {\it\small DGTwoDIS} suffer from a high time consumption when $\beta$ is small, \ie, the number of edges in the graph is huge.
One thing worth noting is that {\it\small DGOneDIS} and {\small\it DGTwoDIS} maintain a solution with the same size all the time.
This is because the power-law graphs are easy to process, so only the degree-one reduction will be applied to the vertices when constructing the dependency graph index.

	\section{Conclusion}\label{sec:con}
In this paper, we develop a framework that efficiently maintains a $k$-maximal independent set over dynamic graphs.
We prove that the maintained result is a $(\frac{\Delta}{2} + 1)$-approximate MaxIS in general graphs and a constant-factor approximate MaxIS in power-law bounded graphs with parameters $\delta = 1$ and $\beta > 2$, which is quite common in real-world networks.
To the best of our knowledge, this is the first work that maintains an approximate MaxIS with non-trivial theoretical accuracy guarantee.
We also give out the lower bound on the approximation ratio achieved by all swap-based algorithms for the MaxIS problem, which indicates the limitation of this methodology.
Following the framework, we instantiate a linear-time dynamic approximation algorithm that maintains an 1-maximal independent set, and a expected near-linear-time dynamic approximation algorithm that maintains a 2-maximal independent set.
Extensive empirical studies demonstrate that the proposed algorithms maintain much larger independent sets while having less running time as the number of update operations increases.
For future directions, there are two possible ways.
On the one hand, a better approximation ratio may be achieved by utilizing other structural information;
on the other hand, applying other optimization strategies to the framework may break the worst case sometimes, which may further improve the quality of the solution in practice.

	\section*{Acknowledgments}
	
	This work is supported by the National Natural Science Foundation of China (NSFC) Grant NOs. 61732003, 61832003, 61972110, U1811461 and U19A2059, and the National Key R\&D Program of China Grant NO. 2019YFB2101900.
	
	\bibliographystyle{IEEEtran}
	\bibliography{DynamicMIS_ref.bib}

% Generated by IEEEtran.bst, version: 1.12 (2007/01/11)
\begin{thebibliography}{10}
\providecommand{\url}[1]{#1}
\csname url@samestyle\endcsname
\providecommand{\newblock}{\relax}
\providecommand{\bibinfo}[2]{#2}
\providecommand{\BIBentrySTDinterwordspacing}{\spaceskip=0pt\relax}
\providecommand{\BIBentryALTinterwordstretchfactor}{4}
\providecommand{\BIBentryALTinterwordspacing}{\spaceskip=\fontdimen2\font plus
\BIBentryALTinterwordstretchfactor\fontdimen3\font minus
  \fontdimen4\font\relax}
\providecommand{\BIBforeignlanguage}[2]{{%
\expandafter\ifx\csname l@#1\endcsname\relax
\typeout{** WARNING: IEEEtran.bst: No hyphenation pattern has been}%
\typeout{** loaded for the language `#1'. Using the pattern for}%
\typeout{** the default language instead.}%
\else
\language=\csname l@#1\endcsname
\fi
#2}}
\providecommand{\BIBdecl}{\relax}
\BIBdecl

\bibitem{DBLP:books/fm/GareyJ79}
M.~R. Garey and D.~S. Johnson, \emph{Computers and Intractability: {A} Guide to
  the Theory of NP-Completeness}.\hskip 1em plus 0.5em minus 0.4em\relax W. H.
  Freeman, 1979.

\bibitem{DBLP:journals/pvldb/FuWCW13}
\BIBentryALTinterwordspacing
A.~W. Fu, H.~Wu, J.~Cheng, and R.~C. Wong, ``{IS-LABEL:} an independent-set
  based labeling scheme for point-to-point distance querying,'' \emph{Proc.
  {VLDB} Endow.}, vol.~6, no.~6, pp. 457--468, 2013. [Online]. Available:
  \url{http://www.vldb.org/pvldb/vol6/p457-fu.pdf}
\BIBentrySTDinterwordspacing

\bibitem{DBLP:journals/pvldb/JiangFWX14}
\BIBentryALTinterwordspacing
M.~Jiang, A.~W. Fu, R.~C. Wong, and Y.~Xu, ``Hop doubling label indexing for
  point-to-point distance querying on scale-free networks,'' \emph{Proc. {VLDB}
  Endow.}, vol.~7, no.~12, pp. 1203--1214, 2014. [Online]. Available:
  \url{http://www.vldb.org/pvldb/vol7/p1203-jiang.pdf}
\BIBentrySTDinterwordspacing

\bibitem{DBLP:journals/jpdc/AraujoFDSK11}
\BIBentryALTinterwordspacing
F.~Ara{\'{u}}jo, J.~Farinha, P.~Domingues, G.~C. Silaghi, and D.~Kondo, ``A
  maximum independent set approach for collusion detection in voting pools,''
  \emph{J. Parallel Distributed Comput.}, vol.~71, no.~10, pp. 1356--1366,
  2011. [Online]. Available: \url{https://doi.org/10.1016/j.jpdc.2011.06.004}
\BIBentrySTDinterwordspacing

\bibitem{DBLP:journals/pvldb/MiaoCLGL20}
\BIBentryALTinterwordspacing
D.~Miao, Z.~Cai, J.~Li, X.~Gao, and X.~Liu, ``The computation of optimal subset
  repairs,'' \emph{Proc. {VLDB} Endow.}, vol.~13, no.~11, pp. 2061--2074, 2020.
  [Online]. Available: \url{http://www.vldb.org/pvldb/vol13/p2061-miao.pdf}
\BIBentrySTDinterwordspacing

\bibitem{DBLP:journals/tcs/MiaoLLL19}
\BIBentryALTinterwordspacing
D.~Miao, X.~Liu, Y.~Li, and J.~Li, ``Vertex cover in conflict graphs,''
  \emph{Theor. Comput. Sci.}, vol. 774, pp. 103--112, 2019. [Online].
  Available: \url{https://doi.org/10.1016/j.tcs.2016.07.009}
\BIBentrySTDinterwordspacing

\bibitem{DBLP:conf/wea/GemsaNR14}
\BIBentryALTinterwordspacing
A.~Gemsa, M.~N{\"{o}}llenburg, and I.~Rutter, ``Evaluation of labeling
  strategies for rotating maps,'' in \emph{Experimental Algorithms - 13th
  International Symposium, {SEA} 2014, Copenhagen, Denmark, June 29 - July 1,
  2014. Proceedings}, ser. Lecture Notes in Computer Science, J.~Gudmundsson
  and J.~Katajainen, Eds., vol. 8504.\hskip 1em plus 0.5em minus 0.4em\relax
  Springer, 2014, pp. 235--246. [Online]. Available:
  \url{https://doi.org/10.1007/978-3-319-07959-2\_20}
\BIBentrySTDinterwordspacing

\bibitem{DBLP:conf/wea/GoldbergHM05}
\BIBentryALTinterwordspacing
M.~K. Goldberg, D.~L. Hollinger, and M.~Magdon{-}Ismail, ``Experimental
  evaluation of the greedy and random algorithms for finding independent sets
  in random graphs,'' in \emph{Experimental and Efficient Algorithms, 4th
  InternationalWorkshop, {WEA} 2005, Santorini Island, Greece, May 10-13, 2005,
  Proceedings}, ser. Lecture Notes in Computer Science, S.~E. Nikoletseas, Ed.,
  vol. 3503.\hskip 1em plus 0.5em minus 0.4em\relax Springer, 2005, pp.
  513--523. [Online]. Available: \url{https://doi.org/10.1007/11427186\_44}
\BIBentrySTDinterwordspacing

\bibitem{DBLP:conf/kdd/ZakiPOL97}
\BIBentryALTinterwordspacing
M.~J. Zaki, S.~Parthasarathy, M.~Ogihara, and W.~Li, ``New algorithms for fast
  discovery of association rules,'' in \emph{Proceedings of the Third
  International Conference on Knowledge Discovery and Data Mining (KDD-97),
  Newport Beach, California, USA, August 14-17, 1997}, D.~Heckerman,
  H.~Mannila, and D.~Pregibon, Eds.\hskip 1em plus 0.5em minus 0.4em\relax
  {AAAI} Press, 1997, pp. 283--286. [Online]. Available:
  \url{http://www.aaai.org/Library/KDD/1997/kdd97-060.php}
\BIBentrySTDinterwordspacing

\bibitem{DBLP:journals/iandc/XiaoN17}
\BIBentryALTinterwordspacing
M.~Xiao and H.~Nagamochi, ``Exact algorithms for maximum independent set,''
  \emph{Inf. Comput.}, vol. 255, pp. 126--146, 2017. [Online]. Available:
  \url{https://doi.org/10.1016/j.ic.2017.06.001}
\BIBentrySTDinterwordspacing

\bibitem{DBLP:journals/jal/Robson86}
\BIBentryALTinterwordspacing
J.~M. Robson, ``Algorithms for maximum independent sets,'' \emph{J.
  Algorithms}, vol.~7, no.~3, pp. 425--440, 1986. [Online]. Available:
  \url{https://doi.org/10.1016/0196-6774(86)90032-5}
\BIBentrySTDinterwordspacing

\bibitem{DBLP:conf/focs/Hastad96}
\BIBentryALTinterwordspacing
J.~H{\aa}stad, ``Clique is hard to approximate within
  n\({}^{\mbox{1-epsilon}}\),'' in \emph{37th Annual Symposium on Foundations
  of Computer Science, {FOCS} '96, Burlington, Vermont, USA, 14-16 October,
  1996}.\hskip 1em plus 0.5em minus 0.4em\relax {IEEE} Computer Society, 1996,
  pp. 627--636. [Online]. Available:
  \url{https://doi.org/10.1109/SFCS.1996.548522}
\BIBentrySTDinterwordspacing

\bibitem{DBLP:journals/siamdm/Feige04}
\BIBentryALTinterwordspacing
U.~Feige, ``Approximating maximum clique by removing subgraphs,'' \emph{{SIAM}
  J. Discret. Math.}, vol.~18, no.~2, pp. 219--225, 2004. [Online]. Available:
  \url{https://doi.org/10.1137/S089548010240415X}
\BIBentrySTDinterwordspacing

\bibitem{DBLP:journals/heuristics/AndradeRW12}
\BIBentryALTinterwordspacing
D.~V. Andrade, M.~G.~C. Resende, and R.~F.~F. Werneck, ``Fast local search for
  the maximum independent set problem,'' \emph{J. Heuristics}, vol.~18, no.~4,
  pp. 525--547, 2012. [Online]. Available:
  \url{https://doi.org/10.1007/s10732-012-9196-4}
\BIBentrySTDinterwordspacing

\bibitem{DBLP:conf/sigmod/ChangLZ17}
\BIBentryALTinterwordspacing
L.~Chang, W.~Li, and W.~Zhang, ``Computing {A} near-maximum independent set in
  linear time by reducing-peeling,'' in \emph{Proceedings of the 2017 {ACM}
  International Conference on Management of Data, {SIGMOD} Conference 2017,
  Chicago, IL, USA, May 14-19, 2017}, S.~Salihoglu, W.~Zhou, R.~Chirkova,
  J.~Yang, and D.~Suciu, Eds.\hskip 1em plus 0.5em minus 0.4em\relax {ACM},
  2017, pp. 1181--1196. [Online]. Available:
  \url{https://doi.org/10.1145/3035918.3035939}
\BIBentrySTDinterwordspacing

\bibitem{DBLP:conf/wea/DahlumLS0SW16}
\BIBentryALTinterwordspacing
J.~Dahlum, S.~Lamm, P.~Sanders, C.~Schulz, D.~Strash, and R.~F. Werneck,
  ``Accelerating local search for the maximum independent set problem,'' in
  \emph{Experimental Algorithms - 15th International Symposium, {SEA} 2016, St.
  Petersburg, Russia, June 5-8, 2016, Proceedings}, ser. Lecture Notes in
  Computer Science, A.~V. Goldberg and A.~S. Kulikov, Eds., vol. 9685.\hskip
  1em plus 0.5em minus 0.4em\relax Springer, 2016, pp. 118--133. [Online].
  Available: \url{https://doi.org/10.1007/978-3-319-38851-9\_9}
\BIBentrySTDinterwordspacing

\bibitem{DBLP:journals/heuristics/GrossoLP08}
\BIBentryALTinterwordspacing
A.~Grosso, M.~Locatelli, and W.~J. Pullan, ``Simple ingredients leading to very
  efficient heuristics for the maximum clique problem,'' \emph{J. Heuristics},
  vol.~14, no.~6, pp. 587--612, 2008. [Online]. Available:
  \url{https://doi.org/10.1007/s10732-007-9055-x}
\BIBentrySTDinterwordspacing

\bibitem{DBLP:conf/alenex/LammS0SW16}
\BIBentryALTinterwordspacing
S.~Lamm, P.~Sanders, C.~Schulz, D.~Strash, and R.~F. Werneck, ``Finding
  near-optimal independent sets at scale,'' in \emph{Proceedings of the
  Eighteenth Workshop on Algorithm Engineering and Experiments, {ALENEX} 2016,
  Arlington, Virginia, USA, January 10, 2016}, M.~T. Goodrich and
  M.~Mitzenmacher, Eds.\hskip 1em plus 0.5em minus 0.4em\relax {SIAM}, 2016,
  pp. 138--150. [Online]. Available:
  \url{https://doi.org/10.1137/1.9781611974317.12}
\BIBentrySTDinterwordspacing

\bibitem{DBLP:journals/pvldb/LiuLYXW15}
\BIBentryALTinterwordspacing
Y.~Liu, J.~Lu, H.~Yang, X.~Xiao, and Z.~Wei, ``Towards maximum independent sets
  on massive graphs,'' \emph{Proc. {VLDB} Endow.}, vol.~8, no.~13, pp.
  2122--2133, 2015. [Online]. Available:
  \url{http://www.vldb.org/pvldb/vol8/p2122-lu.pdf}
\BIBentrySTDinterwordspacing

\bibitem{DBLP:conf/icde/ZhengWYC018}
\BIBentryALTinterwordspacing
W.~Zheng, Q.~Wang, J.~X. Yu, H.~Cheng, and L.~Zou, ``Efficient computation of a
  near-maximum independent set over evolving graphs,'' in \emph{34th {IEEE}
  International Conference on Data Engineering, {ICDE} 2018, Paris, France,
  April 16-19, 2018}.\hskip 1em plus 0.5em minus 0.4em\relax {IEEE} Computer
  Society, 2018, pp. 869--880. [Online]. Available:
  \url{https://doi.org/10.1109/ICDE.2018.00083}
\BIBentrySTDinterwordspacing

\bibitem{DBLP:conf/icde/ZhengPCY19}
\BIBentryALTinterwordspacing
W.~Zheng, C.~Piao, H.~Cheng, and J.~X. Yu, ``Computing a near-maximum
  independent set in dynamic graphs,'' in \emph{35th {IEEE} International
  Conference on Data Engineering, {ICDE} 2019, Macao, China, April 8-11,
  2019}.\hskip 1em plus 0.5em minus 0.4em\relax {IEEE}, 2019, pp. 76--87.
  [Online]. Available: \url{https://doi.org/10.1109/ICDE.2019.00016}
\BIBentrySTDinterwordspacing

\bibitem{DBLP:conf/stoc/AielloCL00}
\BIBentryALTinterwordspacing
W.~Aiello, F.~R.~K. Chung, and L.~Lu, ``A random graph model for massive
  graphs,'' in \emph{Proceedings of the Thirty-Second Annual {ACM} Symposium on
  Theory of Computing, May 21-23, 2000, Portland, OR, {USA}}, F.~F. Yao and
  E.~M. Luks, Eds.\hskip 1em plus 0.5em minus 0.4em\relax {ACM}, 2000, pp.
  171--180. [Online]. Available: \url{https://doi.org/10.1145/335305.335326}
\BIBentrySTDinterwordspacing

\bibitem{DBLP:journals/algorithmica/ChauhanFR20}
\BIBentryALTinterwordspacing
A.~Chauhan, T.~Friedrich, and R.~Rothenberger, ``Greed is good for
  deterministic scale-free networks,'' \emph{Algorithmica}, vol.~82, no.~11,
  pp. 3338--3389, 2020. [Online]. Available:
  \url{https://doi.org/10.1007/s00453-020-00729-z}
\BIBentrySTDinterwordspacing

\bibitem{DBLP:conf/soda/BrachCLS16}
\BIBentryALTinterwordspacing
P.~Brach, M.~Cygan, J.~Lacki, and P.~Sankowski, ``Algorithmic complexity of
  power law networks,'' in \emph{Proceedings of the Twenty-Seventh Annual
  {ACM-SIAM} Symposium on Discrete Algorithms, {SODA} 2016, Arlington, VA, USA,
  January 10-12, 2016}, R.~Krauthgamer, Ed.\hskip 1em plus 0.5em minus
  0.4em\relax {SIAM}, 2016, pp. 1306--1325. [Online]. Available:
  \url{https://doi.org/10.1137/1.9781611974331.ch91}
\BIBentrySTDinterwordspacing

\bibitem{snapnets}
J.~Leskovec and A.~Krevl, ``{SNAP Datasets}: {Stanford} large network dataset
  collection,'' \url{http://snap.stanford.edu/data}, Jun. 2014.

\bibitem{DBLP:books/sp/03/LourencoMS03}
\BIBentryALTinterwordspacing
H.~R. Louren{\c{c}}o, O.~C. Martin, and T.~St{\"{u}}tzle, ``Iterated local
  search,'' in \emph{Handbook of Metaheuristics}, ser. International Series in
  Operations Research {\&} Management Science, F.~W. Glover and G.~A.
  Kochenberger, Eds.\hskip 1em plus 0.5em minus 0.4em\relax Kluwer / Springer,
  2003, vol.~57, pp. 320--353. [Online]. Available:
  \url{https://doi.org/10.1007/0-306-48056-5\_11}
\BIBentrySTDinterwordspacing

\bibitem{BoVWFI}
P.~Boldi and S.~Vigna, ``The {W}eb{G}raph framework {I}: {C}ompression
  techniques,'' in \emph{Proc. of the Thirteenth International World Wide Web
  Conference (WWW 2004)}.\hskip 1em plus 0.5em minus 0.4em\relax Manhattan,
  USA: ACM Press, 2004, pp. 595--601.

\bibitem{BRSLLP}
P.~Boldi, M.~Rosa, M.~Santini, and S.~Vigna, ``Layered label propagation: A
  multiresolution coordinate-free ordering for compressing social networks,''
  in \emph{Proceedings of the 20th international conference on World Wide Web},
  S.~Srinivasan, K.~Ramamritham, A.~Kumar, M.~P. Ravindra, E.~Bertino, and
  R.~Kumar, Eds.\hskip 1em plus 0.5em minus 0.4em\relax ACM Press, 2011, pp.
  587--596.

\bibitem{DBLP:journals/tcs/AkibaI16}
\BIBentryALTinterwordspacing
T.~Akiba and Y.~Iwata, ``Branch-and-reduce exponential/fpt algorithms in
  practice: {A} case study of vertex cover,'' \emph{Theor. Comput. Sci.}, vol.
  609, pp. 211--225, 2016. [Online]. Available:
  \url{https://doi.org/10.1016/j.tcs.2015.09.023}
\BIBentrySTDinterwordspacing

\end{thebibliography}
	
\end{document}